\documentclass[aps,pra,twocolumn,superscriptaddress,showpacs,nofootinbibfloatfix,amsmath,amsfonts,amssymb]{revtex4-2}%
\usepackage{amsmath,amsfonts,amssymb,color}
\usepackage{amsthm}
\usepackage{leftidx}
\usepackage{graphicx}
\usepackage{xcolor}
\usepackage{dcolumn}
\usepackage{bm}
\usepackage{epstopdf}
\usepackage{epsfig}
\usepackage{environ}
\usepackage{pdfcomment}

\usepackage{multirow}
\usepackage{setspace}
\usepackage{color}

\usepackage{float}
\usepackage[T1]{fontenc}
\usepackage[latin9]{inputenc}
\usepackage{setspace}
\usepackage{esint}

\usepackage{wasysym}

\providecommand{\tabularnewline}{\\}

\begin{document}


\title{Floquet M\"obius topological insulators}

\author{Longwen Zhou}
\email{zhoulw13@u.nus.edu}
\affiliation{%
	College of Physics and Optoelectronic Engineering, Ocean University of China, Qingdao, China 266100
}
\affiliation{%
	Key Laboratory of Optics and Optoelectronics, Qingdao, China 266100
}
\affiliation{%
	Engineering Research Center of Advanced Marine Physical Instruments and Equipment of MOE, Qingdao, China 266100
}
\author{Fan Zhang}
\email{These authors contributed equally to this work.}
\affiliation{%
	College of Physics and Optoelectronic Engineering, Ocean University of China, Qingdao, China 266100
}
\author{Jiaxin Pan}
\email{These authors contributed equally to this work.}
\affiliation{%
	College of Physics and Optoelectronic Engineering, Ocean University of China, Qingdao, China 266100
}

\date{\today}

\begin{abstract}
M\"obius topological insulators have dispersive
edge bands with M\"obius twists in momentum space, which are protected
by the combination of chiral and $\mathbb{Z}_{2}$-projective translational
symmetries. In this work, we reveal a unique type of M\"obius topological
insulator, whose edge bands could twist around the quasienergy $\pi$
of a periodically driven system and are thus of Floquet origin. By
applying time-periodic quenches to an experimentally realized M\"obius
insulator model, we obtain interconnected M\"obius edge bands
around zero and $\pi$ quasienergies, which can coexist with a gapped
or gapless bulk. These M\"obius bands are topologically characterized
by a pair of generalized winding numbers, which are integer-quantized
due to an emergent chiral symmetry at a high-symmetry point
in momentum space. Numerical investigations of the quasienergy
and entanglement spectra provide consistent evidence for the presence
of such M\"obius topological phases. A protocol based on the adiabatic
switching of edge-band populations is further introduced to dynamically
characterize the topology of Floquet M\"obius edge bands. Our findings
thus extend the scope of M\"obius topological phases to nonequilibrium
settings and unveil a unique class of M\"obius-twisted topological edge
states without static counterparts.
\end{abstract}

\maketitle

\section{Introduction\label{sec:Int}}

As a prototype of topology matter beyond equilibrium, Floquet
topological phases have attracted continued attention over the past
decades (see Refs.~\cite{FTPRev01,FTPRev02,FTPRev03,FTPRev04,FTPRev05,FTPRev06,FTPRev07,FTPRev08,FTPRev09,FTPRev10}
for reviews). Their intriguing features include phases with large
topological invariants and many topological edge modes due to driving
induced long-range couplings \cite{FloLarge01,FloLarge02,FloLarge03,FloLarge04,FloLarge05,FloLarge06},
anomalous edge states degenerating at quasienergy $\pi$ or coiling
around the whole quasienergy Brillouin zone~\cite{AFTP01,AFTP02,AFTP03,AFTP04,AFTP05,AFTP06,AFTP07,AFTP08},
the recently discovered Floquet phase boundaries with critical $\pi$
edge modes and nontrivial gapless topology \cite{FgSPT01,FgSPT02,FgSPT03},
and so on. These Floquet topological phases have been classified in theory
according to their protecting symmetries \cite{FloClass01,FloClass02,FloClass03},
yielding the periodic table of Floquet topological insulators in analogy
with their static cousins \cite{FloquetPT}.

Beyond the time-reversal, particle-hole and chiral symmetries, other crystal
symmetries may also facilitate topological matter in or out of equilibrium
\cite{FloCry01,FloCry02,FloCry03,FloCry04,FloCry05,FloCry06,FloCry07,FloCry08,FloCry09,FloCry092,FloCry10,FloCry11,FloCry12,FloCry13,FloCry14,FloCry15,FloCry16,FloCry17,FloCry18,FloCry19,FloCry20,FloCry21,FloCry22,FloCry23}.
For example, enriched by nonsymmorphic space groups or projective translational
symmetry (PTS), energy bands may undergo M\"obius
twists in momentum space \cite{MTI0}, yielding
topological phases beyond the conventional tenfold
way \cite{TenfoldNJP}.
Such M\"obius topological phases have attracting increasing interest
\cite{Sym1,MTI01,MTI02,MTI03,MTI04,MTI05,MTI06,MTI07,MTI08,MTI09,MTI10,MTI11,MTI12,MTI13,MTI14,MTI15,MTI16,MTI17,MTI18,MTI19,MTI20,MTI21},
with the M\"obius topological insulator be an
example \cite{MTI07}, whose edge spectra are interconnected like a M\"obius
band under the protection of chiral and $\mathbb{Z}_{2}$-PTS. 
These spectrally interconnected edge bands with a M\"obius twist are referred to as \textit{M\"obius edge bands}.
Besides theoretical progress,
experimental studies of M\"obius topological matter have been conducted
in acoustic \cite{MTI10,MTI11,MTI17}, photonic \cite{MTI14,MTI15,MTI19}
and electrical systems \cite{MTI16}, inspiring potential applications
in the design of wave-guiding devices \cite{MTI22}.

In a static M\"obius insulator, the chiral
symmetry can enforce the degeneracy of edge zero modes at a high-symmetry point in momentum
space, which serves as the twisting point of M\"obius edge bands. 
In the Floquet setting, the chiral symmetry could
allow degenerate edge modes to appear at not only the zero but also the $\pi$
quasienergy.
The latter follows from the fact that a chiral symmetric Floquet operator $U$ satisfies
${\cal S}U{\cal S}=U^{\dagger}$, where ${\cal S}$ is the chiral symmetry operator and ${\cal S}^2=1$.
If $|\psi\rangle$ is an eigenstate of $U$ with quasienergy $\pi$, i.e., 
$U|\psi\rangle=e^{-i\pi}|\psi\rangle$, we must have ${\cal S}U{\cal S}{\cal S}|\psi\rangle=U^{\dagger}{\cal S}|\psi\rangle=e^{-i\pi}{\cal S}|\psi\rangle$,
making ${\cal S}|\psi\rangle$ another eigenstate of $U$ with quasienergy $-\pi$.
As the Floquet spectrum of $U$ is defined mod $2\pi$, the quasienergies $-\pi$ and $\pi$ are identified, and their associated
eigenstates must be degenerate. Therefore, if $|\psi\rangle$ happens to be the edge state of a chiral symmetric $U$ at
quasienergy $\pi$, it must be at least twofold degenerate.
Such degenerated $\pi$ edge modes can be generated via closing and reopening of the Floquet spectral gap at quasienergy $\pi$
during a topological phase transition.
Importantly, this makes it possible to have
a new type of M\"obius edge
band twisted at quasienergy $\pi$ in a chiral symmetric Floquet system when the PTS is also present. Such anomalous Floquet M\"obius edge bands
and their associated topological phases have yet to be revealed.

In this work, we apply time-periodic drivings to an experimentally
realized two-dimensional (2D) M\"obius insulator model \cite{MTI10,MTI11}
and obtain intriguing gapped and gapless M\"obius topological phases
of Floquet origin. Our findings reveal that the collaboration of chiral
symmetry and $\mathbb{Z}_{2}$-PTS could bring about two types of
M\"obius edge bands, with one of them having no equilibrium analogs.
The rest of this paper is organized as follows. In Sec.~\ref{sec:Mod},
we introduce our model from bottom up by stacking one-dimensional
(1D) periodically quenched Su-Schrieffer-Heeger (PQSSH) chains along
a second dimension and applying a $\pi$ magnetic flux to each plaquette
of the resulting 2D lattice. In Sec.~\ref{sec:Res}, we
show that our introduced 2D $\pi$-flux PQSSH model possesses rich
Floquet M\"obius topological insulator phases and gapless critical points,
which are featured by doubly degenerated M\"obius edge bands twisting
around zero and $\pi$ quasienergies. They are further characterized
by a pair of generalized topological winding numbers quantized at
a high-symmetry momentum due to an emergent chiral symmetry.
The correspondence between these winding numbers and the numbers of
Floquet M\"obius edge bands is also established in both gapped and gapless
phases. These findings are then demonstrated numerically by investigating
the quasienergy spectrum, entanglement spectrum, and periodic-doubling
dynamics of M\"obius edge states. Finally, we summarize
our results, discuss the experimental feasibility of our model and
the potential applications of Floquet M\"obius topological states in
Sec.~\ref{sec:Sum}. 
A brief recap of the static Su-Schrieffer-Heeger
(SSH) chain \cite{SSH} is given in Appendix \ref{sec:SSH}. Further
details about the theoretical derivation and experimental proposal are
given in Appendices \ref{sec:symmetry} and \ref{sec:exp}.

To avoid possible confusion in terminology, we use the term \textit{bulk-gapless} for a Floquet system to refer to the case in which the bulk Floquet bands in quasienergy regions $E\in[-\pi,0]$ and $E\in[0,\pi]$ are touched at least at one point in the first Brillouin zone, so that there is no quasienergy gap between any pair of spectrally adjacent bulk states. \textit{Topologically protected edge states} correspond to edge-localized (Floquet) states that are characterized by nonzero topological invariants and protected by relevant symmetries of the underlying model.

\section{Models\label{sec:Mod}}

\begin{figure}
	\begin{centering}
		\includegraphics[scale=0.335]{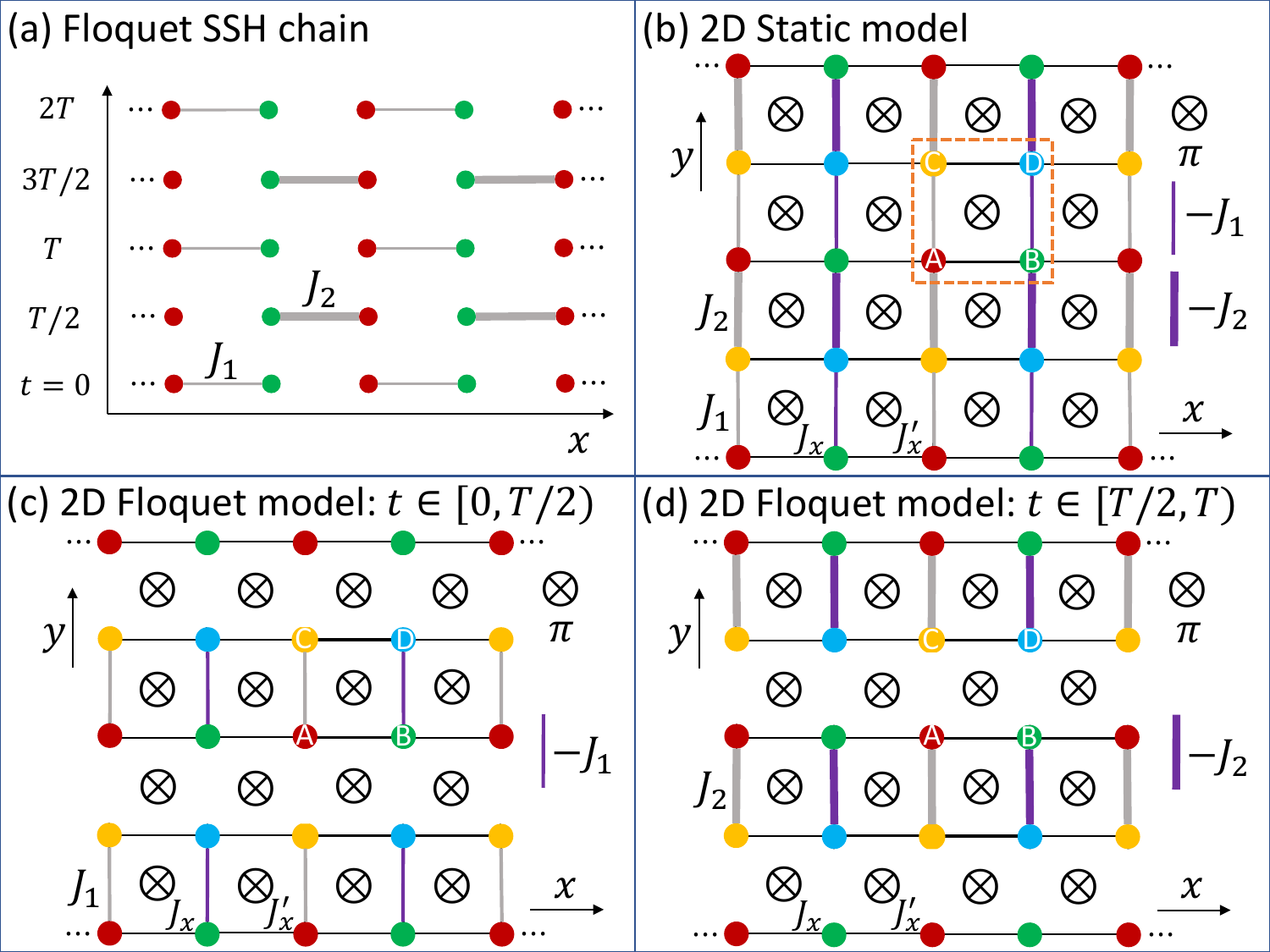}
		\par\end{centering}
	\caption{Illustration of the PQSSH model and its 2D extension. (a) shows an
		SSH chain (lying along the $x$ direction at each time $t$) under time-periodic
		quenches over two driving periods ($t=0\rightarrow2T$). The $J_{1}$
		and $J_{2}$ denote intracell and intercell hopping amplitudes.
		The red and green dots denote sublattices A and B. (b) shows a
		2D extension of the SSH model with a $\pi$ magnetic flux per plaquette.
		The orange dashed line encircles a unit cell with four sublattices
		A, B, C and D. (c) and (d) show the 2D $\pi$-flux PQSSH model in
		its first and second half of a evolution period, respectively.
		We set the intracell and intercell hopping amplitudes along the $x$ direction
		in (b)--(d) to $J_{x}=J'_{x}=J$ throughout the paper. \label{fig:Sketch}}
\end{figure}

In this section, we take a bottom-up approach to build a \emph{minimal}
model of Floquet M\"obius topological insulators (FMTIs). A schematic
illustration of the construction is shown in Fig.~\ref{fig:Sketch}.
We start by introducing a PQSSH chain, which possesses various topological
features that are unique to Floquet systems. Next, we construct our
minimal model via coupling PQSSH chains along a second spatial dimension,
and applying a $\pi$ magnetic flux to each plaquette of the resulting
2D square lattice. A theoretical framework, from the unified perspective
of spectrum, topology, entanglement, and dynamics is developed to
characterize the physical properties of our 2D Floquet system in 
Sec.~\ref{sec:Res}. This framework is applicable to Floquet M\"obius topological
phases in more general situations. Basic topological properties of
the 1D SSH chain are briefly reviewed in the Appendix \ref{sec:SSH}.

\subsection{PQSSH model\label{subsec:PQSSH}}

The SSH model forms a paradigm in the study of topological insulators.
We start our investigation by adding time-periodic drivings to the
SSH model. For simplicity, we consider piecewise quenches applied
to the hopping amplitudes of the system, which are experimentally
realizable in quantum simulators like photonic \cite{AFTP04} and
acoustic \cite{FloCry23} waveguides. The resulting time-dependent
Hamiltonian $\hat{H}(t)$ takes the form
\begin{equation}
	\hat{H}(t)=\begin{cases}
		\hat{H}_{1}, & t\in[\ell T,\ell T+T/2)\\
		\hat{H}_{2}, & t\in[\ell T+T/2,\ell T+T]
	\end{cases},\label{eq:HtSSH}
\end{equation}
where $\ell\in\mathbb{Z}$, $T$ is the driving period, and
\begin{alignat}{1}
	\hat{H}_{1}= & \sum_{j}J_{1}\hat{a}_{j}^{\dagger}\hat{b}_{j}+{\rm H.c.},\label{eq:H1SSH}\\
	\hat{H}_{2}= & \sum_{j}J_{2}\hat{b}_{j}^{\dagger}\hat{a}_{j+1}+{\rm H.c.}.\label{eq:H2SSH}
\end{alignat}
The $\hat{a}_{j}^{\dagger}$ ($\hat{b}_{j}^{\dagger}$) creates a
particle in the sublattice A (B) of the unit cell $j$. The intracell
(intercell) hopping amplitude $J_{1}$ ($J_{2}$) is switched on only
within the first (second) half of each driving period $t\in[\ell T,\ell T+T)$.
An illustration of this driven lattice model is shown in Fig.~\ref{fig:Sketch}(a).
The Floquet operator of the system, which controls its evolution over
a full driving period $T$ reads
\begin{equation}
	\hat{U}=e^{-i\frac{T}{2\hbar}\hat{H}_{2}}e^{-i\frac{T}{2\hbar}\hat{H}_{1}}.\label{eq:USSH}
\end{equation}

From now on, we set $2\hbar/T=1$ as the unit of energy. Under
the periodic boundary condition (PBC), we can express the $\hat{U}$
in momentum space as $\hat{U}=\sum_{k}\hat{\Psi}_{k}^{\dagger}U(k)\hat{\Psi}_{k}$,
where 
\begin{alignat}{1}
	U(k)& = e^{-ih_{2}(k)}e^{-ih_{1}(k)},\label{eq:UkSSH}\\
	h_{1}(k)& = J_{1}\sigma_{x},\label{eq:h1kSSH}\\
	h_{2}(k)& = J_{2}(\cos k\sigma_{x}+\sin k\sigma_{y}).\label{eq:h2kSSH}
\end{alignat}
$\sigma_x$ and $\sigma_y$ are Pauli matrices. The Floquet quasienergy dispersion can be obtained by solving the
eigenvalue equation $U(k)|\psi(k)\rangle=e^{-iE(k)}|\psi(k)\rangle$,
yielding 
\begin{equation}
	\cos[E(k)]=\cos J_{1}\cos J_{2}-\sin J_{1}\sin J_{2}\cos k.\label{eq:QEkSSH}
\end{equation}
There are two Floquet bands with quasienergies $E_{\pm}(k)=\pm E(k)=\pm\arccos(\cos J_{1}\cos J_{2}-\sin J_{1}\sin J_{2}\cos k)$.
Different from the static case (see Appendix \ref{sec:SSH} for details),
these Floquet bands could meet with each other at either $E=0$ or
$E=\pm\pi$, yielding two possible flavors of phase transitions. Referring
to the Eq.~(\ref{eq:QEkSSH}), these transitions could happen only
if $\cos[E(k)]=\pm1$, yielding the following phase boundary equations
in the $J_{1}$-$J_{2}$ plane
\begin{equation}
	J_{2}=\nu\pi\pm J_{1},\quad\nu\in\mathbb{Z}.\label{eq:PBSSH}
\end{equation}
The gap between quasienergy bands $E_{\pm}(k)$ closes at $E=0$ ($E=\pm\pi$)
if $\nu\in2\mathbb{Z}$ ($2\mathbb{Z}-1$). To distinguish and characterize
different Floquet phases that are separated by these phase boundaries,
we need to first identify the symmetries of the driven system. This
is achieved by transforming the Floquet operator $U(k)$ to a pair
of symmetric time frames \cite{STF01,STF02,STF03}, yielding
\begin{alignat}{1}
	U_{1}(k)& = e^{-\frac{i}{2}h_{2}(k)}e^{-ih_{1}(k)}e^{-\frac{i}{2}h_{2}(k)},\label{eq:U1kSSH}\\
	U_{2}(k)& = e^{-\frac{i}{2}h_{1}(k)}e^{-ih_{2}(k)}e^{-\frac{i}{2}h_{1}(k)}.\label{eq:U2kSSH}
\end{alignat}
As the $U_{1,2}(k)$ are related to the $U(k)$ in Eq.~(\ref{eq:UkSSH})
via unitary transformations, they share the same Floquet spectrum
$E_{\pm}(k)$. Meanwhile, the $U_{s}(k)$ ($s=1,2$) has the chiral
symmetry ${\cal S}=\sigma_{z}$ with ${\cal S}U_{s}(k){\cal S}=U_{s}^{\dagger}(k)$
and ${\cal S}^{2}=\sigma_{0}$, the time-reversal symmetry ${\cal T}=\sigma_{0}$
with ${\cal T}U_{s}^{*}(k){\cal T}^{\dagger}=U_{s}^{\dagger}(-k)$
and ${\cal T}^{2}=\sigma_{0}$, the particle-hole symmetry ${\cal C}=\sigma_{z}$
with ${\cal C}U_{s}^{*}(k){\cal C}^{\dagger}=U(-k)$ and ${\cal C}^{2}=\sigma_{0}$,
and the inversion symmetry ${\cal I}=\sigma_{x}$ with ${\cal I}U(k){\cal I}^{\dagger}=U(-k)$.
It belongs to the BDI symmetry class \cite{STF01,STF02,STF03}.
The gapped Floquet phases of $U(k)$ can be characterized
by a pair of integer topological winding numbers $(w_{0},w_{\pi})$, defined as
\begin{equation}
	(w_{0},w_{\pi})=\frac{1}{2}(w_{1}+w_{2},w_{1}-w_{2}),\label{eq:w0p}
\end{equation}
where 
\begin{equation}
	w_{s}\equiv\int_{-\pi}^{\pi}\frac{dk}{2\pi}\partial_{k}\phi_{s}(k),\label{eq:w12}
\end{equation}
$\phi_{s}(k)=\arctan[h_{sy}(k)/h_{sx}(k)]$, $h_{sx}(k)=\frac{i}{2}{\rm Tr}[\sigma_{x}U_{s}(k)]$,
$h_{sy}(k)=\frac{i}{2}{\rm Tr}[\sigma_{y}U_{s}(k)]$ and $s=1,2$
\cite{STF01,STF02,STF03}.

\begin{figure}
	\begin{centering}
		\includegraphics[scale=0.48]{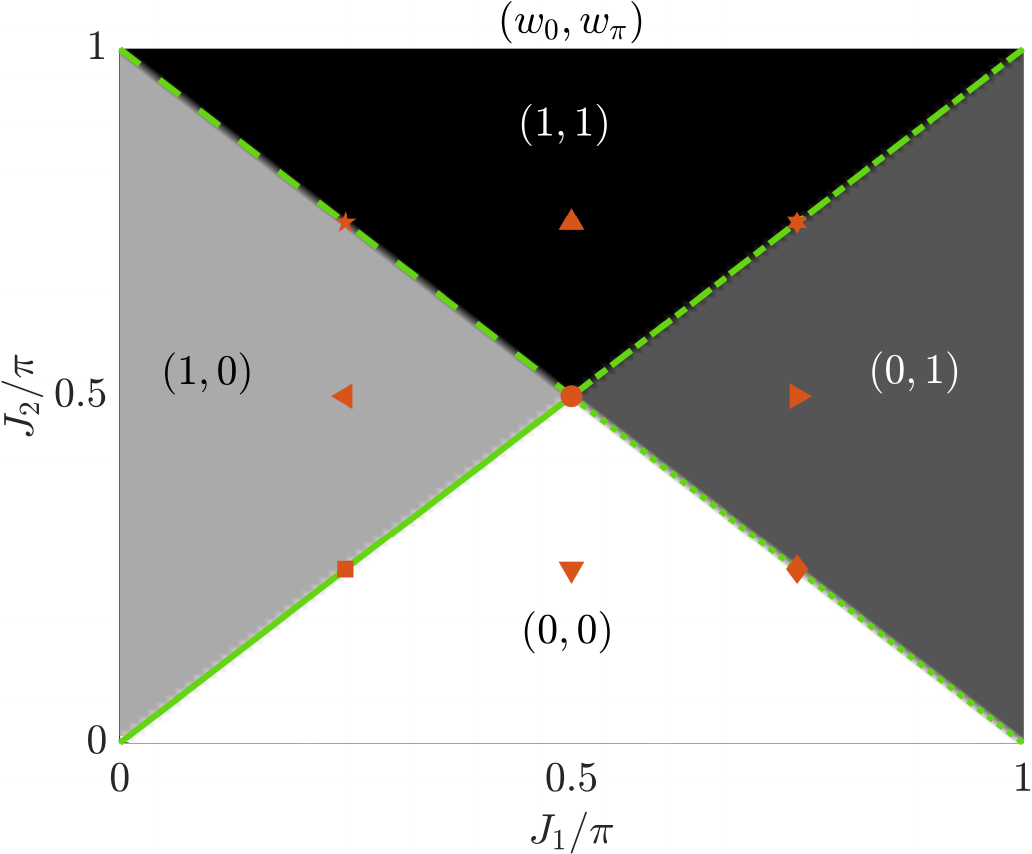}
		\par\end{centering}
	\caption{Topological phase diagram of the PQSSH model, with winding numbers
		$(w_{0},w_{\pi})$ denoted explicitly in each gapped phase. The nine
		data sets used in the calculation of Fig.~\ref{fig:PQSSH2} are highlighted
		by different symbols in the phase diagram. The solid and dotted lines
		are trivial critical lines, with quasienergy gap closes at $E=0$
		and $E=\pi$, respectively. The dashed and dash-dotted lines are topological
		critical lines, with quasienergy gap closes at $E=\pi$ and $E=0$,
		respectively. \label{fig:PQSSH1}}
\end{figure}

\begin{figure*}
	\begin{centering}
		\includegraphics[scale=0.7]{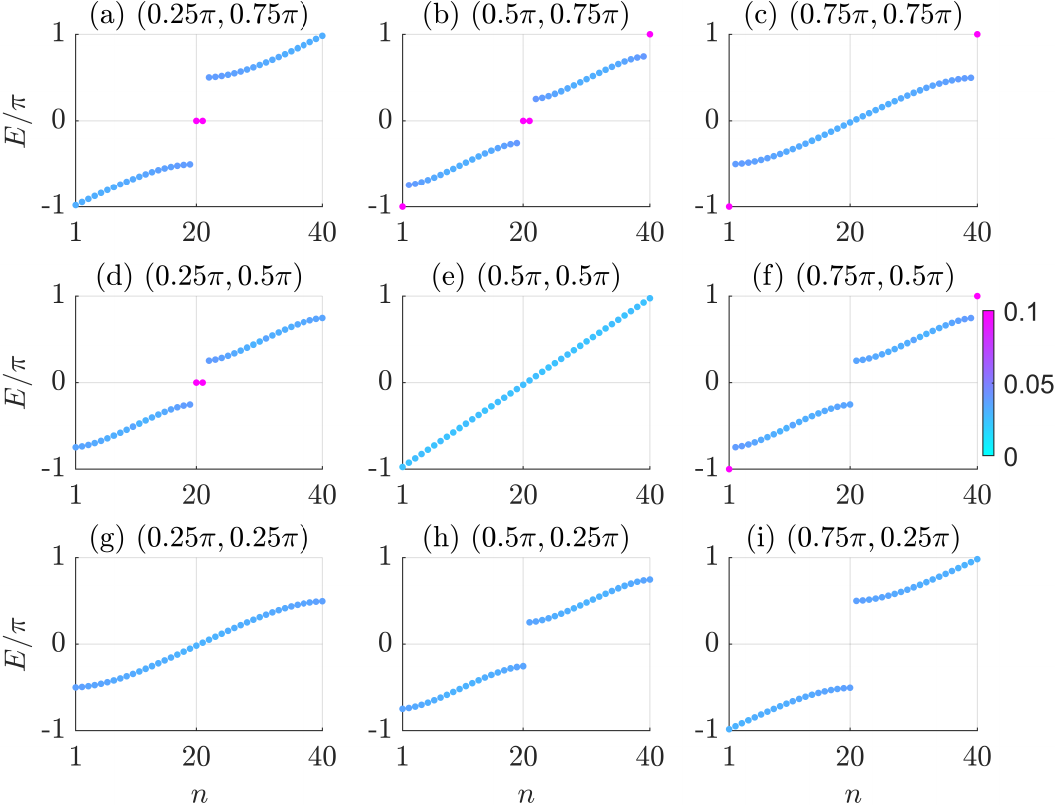}
		\par\end{centering}
	\caption{Floquet spectrum of the PQSSH model under the OBC.
		The values of $(J_{1},J_{2})$ are given in the caption of each panel.
		$20$ unit cells are used in each calculation. $n$ is the state index.
		The color of each data point is given by the inverse participation
		ratio of the corresponding state. The cases in (b), (d) and (f) correspond
		to Floquet topological insulator phases, with system parameters taken
		at the $\blacktriangle$, $\blacktriangleleft$ and $\blacktriangleright$
		in Fig.~\ref{fig:PQSSH1}. The case in (h) corresponds to a trivial
		insulator phase, with system parameters taken at the $\blacktriangledown$
		in Fig.~\ref{fig:PQSSH1}. The cases in (a) and (c) correspond to
		topologically nontrivial critical points, with system parameters taken
		at the $\bigstar$ and six-pointed star in Fig.~\ref{fig:PQSSH1}.
		The cases in (e), (g) and (i) correspond to topologically trivial
		critical points, with system parameters taken at the $\CIRCLE$, $\blacksquare$
		and $\blacklozenge$ in Fig.~\ref{fig:PQSSH1}. \label{fig:PQSSH2}}
\end{figure*}

In Fig.~\ref{fig:PQSSH1}, we show the bulk phase diagram of the PQSSH
model by evaluating the $(w_{0},w_{\pi})$ in Eq.~(\ref{eq:w0p}). We observe that there
are four distinct gapped phases, with one trivial insulator {[}$(w_{0},w_{\pi})=(0,0)${]}
are three Floquet topological insulators. They are separated by two
critical lines along $J_{2}=J_{1}$ and $J_{2}=\pi-J_{1}$, which can
be further classified into four groups of topologically different
critical points \cite{FgSPT01}. To see this, we introduce the Floquet
effective Hamiltonian of $U_{s}(k)$ ($s=1,2$) \cite{STF01} as
\begin{equation}
	H_{s}(k)\equiv\frac{U_{s}^{\dagger}(k)-U_{s}(k)}{2i}=\begin{bmatrix}0 & f_{s}^{*}(k)\\
		f_{s}(k) & 0
	\end{bmatrix}.\label{eq:Hsk}
\end{equation}
The lower off-diagonal element of $H_{s}(k)$ can be extended to the
whole complex plane, yielding the complex continuation function $f_{s}(z)\equiv f_{s}(k\rightarrow-i\ln z)$.
Counting the difference between the zeros and poles of $f_{s}(z)$
inside the unit circle generates a topological integer $\omega_{s}$
($s=1,2$). The combination of $\omega_{1}$ and $\omega_{2}$ further
yields the generalized winding numbers $(\omega_{0},\omega_{\pi})$
as
\begin{equation}
	(\omega_{0},\omega_{\pi})=\frac{1}{2}(\omega_{1}+\omega_{2},\omega_{1}-\omega_{2}).\label{eq:ome0p}
\end{equation}
It has been shown that these invariants are both integer quantized for chiral symmetric driven systems
\cite{FgSPT01}. They satisfy $(\omega_{0},\omega_{\pi})=(w_{0},w_{\pi})$
throughout the gapped phases of the PQSSH model. Moreover, along the
critical line $J_{2}=J_{1}$, we have $(\omega_{0},\omega_{\pi})=(0,0)$
and $(\omega_{0},\omega_{\pi})=(0,1)$ when $J_{1}\in(0,\pi/2]$ and
$J_{1}\in(\pi/2,\pi)$ in Fig.~\ref{fig:PQSSH1}, respectively. These
two segments of phase boundaries are thus topologically distinct and
separated by a ``\emph{phase transition of phase transition}'' at
$J_{1}=J_{2}=\pi/2$, where the spectrum gap of $U(k)$ closes and
reopens at the quasienergy $\pi$ while remaining closed at the quasienergy
zero. Along the critical lines $J_{2}=\pi-J_{1}$, we instead have
$(\omega_{0},\omega_{\pi})=(1,0)$ and $(\omega_{0},\omega_{\pi})=(0,0)$
when $J_{1}\in(0,\pi/2)$ and $J_{1}\in[\pi/2,\pi)$ in Fig.~\ref{fig:PQSSH1},
respectively. These two segments of phase boundaries are thus also topologically
distinct and separated by a ``\emph{phase transition of phase transition}''
at $J_{1}=J_{2}=\pi/2$, where the spectrum gap of $U(k)$ closes
and reopens at the quasienergy zero while remaining closed at the
quasienergy $\pi$. Therefore, despite the gapped phases, we have
two gapless phase boundaries with nontrivial topology, as highlighted
by the dashed and dash-dotted lines Fig.~\ref{fig:PQSSH1}.

The bulk topology of the PQSSH model could manifest as degenerate
zero and $\pi$ Floquet edge modes at the system boundaries under
the open boundary condition (OBC). To see this, we diagonalize the
Floquet operator $\hat{U}$ in Eq.~(\ref{eq:USSH}) numerically in
the lattice representation and show its quasienergy spectrum for typical
cases in Fig.~\ref{fig:PQSSH2}. We find that degenerate zero and
$\pi$ edge modes could appear in both gapped Floquet phases {[}in
Figs.~\ref{fig:PQSSH2}(b), \ref{fig:PQSSH2}(d) and \ref{fig:PQSSH2}(f){]}
and at gapless critical points {[}in Figs.~\ref{fig:PQSSH2}(a) and
\ref{fig:PQSSH2}(c){]}. The conditions for these edge modes to appear
can be obtained by solving the zero and $\pi$ quasienergy solutions
of $\hat{U}$ exactly in thermodynamic limit. For example, if we consider
a half-infinite Floquet chain with the OBC taken at its left edge,
the zero and $\pi$ eigenmodes of $\hat{U}$ are given by \cite{FgSPT01,FgSPT03}
\begin{alignat}{1}
	|\varphi_{{\rm L}}^{(0)}\rangle= & \sum_{j=1}^{\infty}\left[-\frac{\tan(J_{1}/2)}{\tan(J_{2}/2)}\right]^{j-1}\label{eq:0PQSSH}\\
	\times & \left[\cos(J_{1}/2)\hat{a}_{j}^{\dagger}-i\sin(J_{1}/2)\hat{b}_{j}^{\dagger}\right],\nonumber \\
	|\varphi_{{\rm L}}^{(\pi)}\rangle= & \sum_{j=1}^{\infty}\left[-\frac{1}{\tan(J_{1}/2)\tan(J_{2}/2)}\right]^{j-1}\label{eq:pPQSSH}\\
	\times & \left[\sin(J_{1}/2)\hat{a}_{j}^{\dagger}+i\cos(J_{1}/2)\hat{b}_{j}^{\dagger}\right].\nonumber 
\end{alignat}
Each of these solutions has a degenerate partner (due to the chiral
symmetry) if the OBC is also taken at the right end. In the parameter
space of Fig.~\ref{fig:PQSSH1}, we notice that the zero mode $|\varphi_{{\rm L}}^{(0)}\rangle$
exists as a left edge mode if and only if $|\tan(J_{2}/2)|>|\tan(J_{1}/2)|$,
including the two gapped phases with $(\omega_{0},\omega_{\pi})=(1,0)$,
$(1,1)$ and the gapless phase boundary (green dashed line) $J_{2}=\pi-J_{1}$
with $J_{1}\in(0,\pi/2)$. On the other hand, the $\pi$ mode $|\varphi_{{\rm L}}^{(\pi)}\rangle$
represents a left edge mode if and only if $|\tan(J_{1}/2)\tan(J_{2}/2)|>1$,
including the two gapped phases with $(\omega_{0},\omega_{\pi})=(0,1)$,
$(1,1)$ and the gapless phase boundary (green dash-dotted line) $J_{2}=J_{1}$
with $J_{1}\in(\pi/2,\pi)$. Other parameter domains of the Fig.~\ref{fig:PQSSH1},
including the phase boundaries $J_{2}=J_{1}$ with $J_{1}\in(0,\pi/2]$
(green solid line), $J_{2}=\pi-J_{1}$ with $J_{1}\in(\pi/2,\pi)$
(green dotted line), and the gapped phase with $(\omega_{0},\omega_{\pi})=(0,0)$
have neither zero nor $\pi$ Floquet edge modes. These observations
allow us to establish the rule of correspondence between the bulk topological
invariants $(\omega_{0},\omega_{\pi})$ under the PBC and the numbers
of Floquet zero and $\pi$ edge modes $(N_{0},N_{\pi})$ under the
OBC \cite{FgSPT01}, i.e.,
\begin{equation}
	(N_{0},N_{\pi})=2(|\omega_{0}|,|\omega_{\pi}|).\label{eq:BBCPQSSH}
\end{equation}
This relation holds true in both the Floquet insulator phases and
along the gapless phase boundaries of the PQSSH model. It is also
applicable to describe the bulk-edge correspondence of any other 1D,
two-band chiral symmetric driven systems \cite{FgSPT01}.

In summary, we find that a simple periodic driving scheme not only
endows the SSH model with richer topological phases and phase transitions,
but also creates gapped and gapless Floquet topological matter without
static counterparts. These intriguing possibilities allow us to further
obtain Floquet topological phases with unique M\"obius edge bands via
synthesizing static and Floquet SSH models in two dimensions, as detailed
in the following sections.

\subsection{2D PQSSH model with $\mathbb{Z}_{2}$-PTS\label{subsec:PQSSH2D}}
The static and periodically quenched SSH chains constitute ``minimal'' models
to investigate 1D topological phases in and out of equilibrium. We
now group them together to construct a ``minimal'' setup that could
realize Floquet M\"obius topological phases in two dimensions.

We start with a 2D extension of the SSH model, whose geometry is illustrated
in Fig.~\ref{fig:Sketch}(b). Despite dimerized hoppings along both
the $x$ and $y$ directions, there is also a $\pi$ magnetic flux
through each plaquette of the lattice. This model is originally introduced
in the study of higher-order topological insulators \cite{BBH}. In
an appropriate gauge choice, one can associate the $\pi$-flux of
the $\mathbb{Z}_{2}$ gauge field to each hopping amplitude along
the $y$ direction at every other site of the $x$ direction, as highlighted
by the narrow and wide purple bonds in Fig.~\ref{fig:Sketch}(b).
This model can be further simplified by letting the hopping amplitudes
along the $x$ direction to be uniform, i.e., $J_{x}=J'_{x}=J$ in
Fig.~\ref{fig:Sketch}(b). The lattice Hamiltonian of the resulting
system is given by 
\begin{equation}
	\hat{{\cal H}}=\hat{{\cal H}}_{0}+\hat{{\cal H}}_{1}+\hat{{\cal H}}_{2},\label{eq:HBBH}
\end{equation}
where 
\begin{alignat}{1}
	\hat{{\cal H}}_{0}& = J\sum_{m,n}(\hat{a}_{m,n}^{\dagger}\hat{b}_{m,n}+\hat{b}_{m,n}^{\dagger}\hat{a}_{m+1,n}+{\rm H.c.})\nonumber \\
	& + J\sum_{m,n}(\hat{c}_{m,n}^{\dagger}\hat{d}_{m,n}+\hat{d}_{m,n}^{\dagger}\hat{c}_{m+1,n}+{\rm H.c.}),\label{eq:H0}\\
	\hat{{\cal H}}_{1}& =J_{1}\sum_{m,n}(\hat{a}_{m,n}^{\dagger}\hat{c}_{m,n}-\hat{b}_{m,n}^{\dagger}\hat{d}_{m,n}+{\rm H.c.}),\label{eq:H1}\\
	\hat{{\cal H}}_{2}& =J_{2}\sum_{m,n}(\hat{c}_{m,n}^{\dagger}\hat{a}_{m,n+1}-\hat{d}_{m,n}^{\dagger}\hat{b}_{m,n+1}+{\rm H.c.}).\label{eq:H2}
\end{alignat}
The indices $(m,n)\in\mathbb{Z}\times\mathbb{Z}$ are the $x$ and
$y$ coordinates of the unit cell $(m,n)$. The operators $\hat{a}_{m,n}^{\dagger}$,
$\hat{b}_{m,n}^{\dagger}$, $\hat{c}_{m,n}^{\dagger}$ and $\hat{d}_{m,n}^{\dagger}$
create a fermion in the sublattices A, B, C, and D of the the unit
cell $(m,n)$, respectively {[}see Fig.~\ref{fig:Sketch}(b){]}. The
system described by Eq.~(\ref{eq:HBBH}) possesses the PTS and chiral
symmetry, making it possible to hold M\"obius topological insulator
and semimetal phases \cite{MTI07}. Besides the theoretical interest,
this model has also been realized in acoustic crystals \cite{MTI10,MTI11},
photonic waveguides \cite{MTI14,MTI15} and electrical circuits \cite{MTI16,MTI20}.
In the rest of the paper, we manage to show that time-periodic drivings
could greatly enrich the topology of this prototypical static model
and induce M\"obius topological edge bands that are unique to Floquet
systems.

In analogy with the 1D PQSSH model, we consider a piecewise quenching
protocol, in which the intracell (intercell) hopping amplitudes $\pm J_{1}$
($\pm J_{2}$) are switched on only in the first (second) half of
each driving period. The time-dependent Hamiltonian of the system
then takes the form
\begin{equation}
	\hat{{\cal H}}(t)=\begin{cases}
		\hat{{\cal H}}_{0}+\hat{{\cal H}}_{1}, & t\in[\ell T,\ell T+T/2)\\
		\hat{{\cal H}}_{0}+\hat{{\cal H}}_{2}, & t\in[\ell T+T/2,\ell T+T)
	\end{cases}.\label{eq:calHt}
\end{equation}
A graphical illustration of the driven lattice model is shown in 
Figs.~\ref{fig:Sketch}(c)--\ref{fig:Sketch}(d). The Floquet operator
of the system that describing its evolution over a whole driving period
(e.g., from $t=\ell T$ to $\ell T+T^{-}$) is given by
\begin{equation}
	\hat{{\cal U}}=e^{-i(\hat{{\cal H}}_{0}+\hat{{\cal H}}_{2})}e^{-i(\hat{{\cal H}}_{0}+\hat{{\cal H}}_{1})},\label{eq:calU}
\end{equation}
where we have set $2\hbar/T=1$ as the unit of energy. To identify
the key symmetries relevant to the M\"obius topology of Floquet bands,
we need to transform the operator $\hat{{\cal U}}$ from position
to momentum representations. Taking the Fourier transformation $\hat{f}_{m,n}=\frac{1}{\sqrt{N_{x}N_{y}}}\sum_{{\bf k}\in{\rm BZ}}e^{i(k_{x}m+k_{y}n)}\hat{f}_{{\bf k}}$
for $f=a,b,c,d$ and ${\bf k}=(k_{x},k_{y})$, we find
\begin{equation}
	\hat{{\cal H}}_{\alpha}=\sum_{{\bf k}\in{\rm BZ}}\hat{\Psi}_{{\bf k}}^{\dagger}{\cal H}_{\alpha}({\bf k})\hat{\Psi}_{{\bf k}},\quad\alpha=0,1,2,\label{eq:Ha}
\end{equation}
where $N_{x}$ and $N_{y}$ are numbers of unit cells along the $x$
and $y$ directions. $k_{x}$ and $k_{y}$ are the quasimomenta along
the two spatial dimensions, and $\hat{\Psi}_{{\bf k}}^{\dagger}\equiv(\hat{a}_{{\bf k}}^{\dagger},\hat{b}_{{\bf k}}^{\dagger},\hat{c}_{{\bf k}}^{\dagger},\hat{d}_{{\bf k}}^{\dagger})$.
The Hamiltonian components have the explicit expressions
\begin{alignat}{1}
	{\cal H}_{0}({\bf k})& = J\sigma_{0}\otimes[(1+\cos k_{x})\sigma_{x}+\sin k_{x}\sigma_{y}],\label{eq:H0k}\\
	{\cal H}_{1}({\bf k})& = J_{1}\sigma_{x}\otimes\sigma_{z},\label{eq:H1k}\\
	{\cal H}_{2}({\bf k})& = J_{2}(\cos k_{y}\sigma_{x}+\sin k_{y}\sigma_{y})\otimes\sigma_{z},\label{eq:H2k}
\end{alignat}
where $\sigma_0$ and $\sigma_z$ are the $2\times2$ identity and Pauli matrices. The Floquet operator of the system now takes the form $\hat{{\cal U}}=\sum_{{\bf k}\in{\rm BZ}}\hat{\Psi}_{{\bf k}}^{\dagger}{\cal U}({\bf k})\hat{\Psi}_{{\bf k}}$,
where 
\begin{alignat}{1}
	{\cal U}({\bf k})& = e^{-i{\cal H}_{02}({\bf k})}e^{-i{\cal H}_{01}({\bf k})},\label{eq:Uk}\\
	{\cal H}_{01}({\bf k})& \equiv {\cal H}_{0}({\bf k})+{\cal H}_{1}({\bf k}),\label{eq:calH01k}\\
	{\cal H}_{02}({\bf k})& \equiv {\cal H}_{0}({\bf k})+{\cal H}_{2}({\bf k}).\label{eq:calH02k}
\end{alignat}
In the next section, we analyze the topological properties of the
system described by the $\hat{{\cal U}}$ in Eq.~(\ref{eq:calU}), which
will be referred to as the 2D $\pi$-flux PQSSH model. We will see
that the topological characterization of the 1D PQSSH chain, as discussed
in Sec.~\ref{subsec:PQSSH}, also plays a key role in understanding
the topological origin of Floquet M\"obius edge bands in our 2D driven
system and their relations to the bulk topological indices.

\section{Theory and Results\label{sec:Res}}
In this section, we first develop the bulk theoretical characterization
of Floquet M\"obius topological phases in Sec.~\ref{subsec:BMT}. Different
types of Floquet M\"obius edge bands and their correspondence to the
bulk topological invariants are then established in Sec.~\ref{subsec:FMEB}.
Further evidence for the presence of Floquet M\"obius edge bands and
their topological nature are supplied by investigating the entanglement
spectrum and the adiabatic pumping dynamics of edge-state populations in 
Sec.~\ref{subsec:EAD}.

\subsection{Bulk M\"obius topology\label{subsec:BMT}}
To reveal the bulk topological properties of our 2D $\pi$-flux PQSSH
model, we need to first identify the symmetries of its Floquet operator
$\hat{{\cal U}}$ in Eq.~(\ref{eq:calU}). In momentum space, the
terms ${\cal H}_{\alpha}({\bf k})$ for $\alpha=0,1,2$ in Eqs.~(\ref{eq:H0k})--(\ref{eq:H2k})
and their static combination ${\cal H}({\bf k})={\cal H}_{0}({\bf k})+{\cal H}_{1}({\bf k})+{\cal H}_{2}({\bf k})$
have a set of spatial and internal symmetries, as summarized in the
first to third columns of Table \ref{tab:1}. Whenever the hopping
amplitudes along $y$ satisfy $J_{1}\neq J_{2}$, the primitive translational
symmetry $\mathsf{L}_{y}=\begin{pmatrix}0 & 1\\
	e^{ik_{y}} & 0
\end{pmatrix}\otimes\sigma_{0}$ of ${\cal H}({\bf k})$ is broken. The chiral symmetry $\mathsf{S}$
and the PTS $\mathsf{L}_{x}$ then allow the static system ${\cal H}({\bf k})$
to possess interconnected topological edge bands with a M\"obius twist
along $k_{x}$ when the OBC is taken along the $y$ direction, yielding
a static M\"obius topological insulator protected by the symmetries
$\mathsf{S}$ and $\mathsf{L}_{x}$ \cite{MTI07}. However, these
symmetries cannot be directly carried over to the Floquet operator
${\cal U}({\bf k})$ in Eq.~(\ref{eq:Uk}). For example, for the chiral
symmetry $\mathsf{S}$, we have $\mathsf{S}{\cal U}({\bf k})\mathsf{S}=e^{i[{\cal H}_{0}({\bf k})+{\cal H}_{2}({\bf k})]}e^{i[{\cal H}_{0}({\bf k})+{\cal H}_{1}({\bf k})]}\neq{\cal U}^{\dagger}({\bf k})$
as long as $[{\cal H}_{1}({\bf k}),{\cal H}_{2}({\bf k})]\neq0$.
This issue can be resolved by working in symmetric time frames in
parallel with the case of the 1D PQSSH model \cite{STF01,STF02,STF03}.
Specially, we obtain symmetrized Floquet operators after unitary transformations of the Eq.~(\ref{eq:Uk})
as
\begin{alignat}{1}
	{\cal U}_{1}({\bf k})& = e^{-\frac{i}{2}{\cal H}_{02}({\bf k})}e^{-i{\cal H}_{01}({\bf k})}e^{-\frac{i}{2}{\cal H}_{02}({\bf k})},\label{eq:calU1k}\\
	{\cal U}_{2}({\bf k})& = e^{-\frac{i}{2}{\cal H}_{01}({\bf k})}e^{-i{\cal H}_{02}({\bf k})}e^{-\frac{i}{2}{\cal H}_{01}({\bf k})}.\label{eq:calU2k}
\end{alignat}
The operators ${\cal U}_{1,2}({\bf k})$ are unitary equivalent to
the ${\cal U}({\bf k})$ in Eq.~(\ref{eq:Uk}), implying that they
share the same Floquet spectrum. Meanwhile, both the ${\cal U}_{1}({\bf k})$
and ${\cal U}_{2}({\bf k})$ have the same set of spatial and internal
symmetries as the static Hamiltonian ${\cal H}({\bf k})$, as summarized
in the first, second and fourth columns of Table \ref{tab:1}. According
to the periodic table of Floquet topological insulators \cite{FloquetPT},
the 2D $\pi$-flux PQSSH model belongs to the symmetry class BDI, whose
first-order topological phases are all trivial in two spatial dimensions.
However, the PTS $\mathsf{L}_{x}$ along the $x$ direction, in collaboration with the chiral symmetry $\mathsf{S}$, allow the
system to exhibit Floquet topological edge bands with M\"obius twists,
as discussed below.

\begin{table*}
	\caption{Symmetries of the static Hamiltonian ${\cal H}({\bf k})={\cal H}_{0}({\bf k})+{\cal H}_{1}({\bf k})+{\cal H}_{2}({\bf k})$
		in terms of its components ${\cal H}_{\alpha}({\bf k})$ ($\alpha=0,1,2$)
		in Eqs.~(\ref{eq:H0k})--(\ref{eq:H2k}), and the Floquet operator
		${\cal U}_{s}({\bf k})$ ($s=1,2$) in Eqs.~(\ref{eq:calU1k}) and
		(\ref{eq:calU2k}). \label{tab:1}}
	
	\centering{}%
	\begin{tabular}{c|c|c|c}
		\hline 
		\textbf{Symmetry} & \textbf{Representation} & \textbf{Static Hamiltonian} & \textbf{Floquet Operator}\tabularnewline
		\hline 
		\hline 
		Time-Reversal & $\mathsf{T}=\sigma_{0}\otimes\sigma_{0}$ & $\mathsf{T}{\cal H}_{\alpha}^{*}({\bf k})\mathsf{T}^{\dagger}={\cal H}_{\alpha}(-{\bf k})$ & $\mathsf{T}{\cal U}_{s}^{*}({\bf k})\mathsf{T}^{\dagger}={\cal U}_{s}^{\dagger}(-{\bf k})$\tabularnewline
		\hline 
		Particle-Hole & $\mathsf{C}=\sigma_{z}\otimes\sigma_{z}$ & $\mathsf{C}{\cal H}_{\alpha}^{*}({\bf k})\mathsf{C}^{\dagger}=-{\cal H}_{\alpha}(-{\bf k})$ & $\mathsf{C}{\cal U}_{s}^{*}({\bf k})\mathsf{C}^{\dagger}={\cal U}_{s}(-{\bf k})$\tabularnewline
		\hline 
		Chiral & $\mathsf{S}=\sigma_{z}\otimes\sigma_{z}$ & $\mathsf{S}{\cal H}_{\alpha}({\bf k})\mathsf{S}=-{\cal H}_{\alpha}({\bf k})$ & $\mathsf{S}{\cal U}_{s}({\bf k})\mathsf{S}={\cal U}_{s}^{\dagger}({\bf k})$\tabularnewline
		\hline 
		Inversion & $\mathsf{I}=\sigma_{y}\otimes\sigma_{x}$ & $\mathsf{I}{\cal H}_{\alpha}({\bf k})\mathsf{I}^{\dagger}={\cal H}_{\alpha}(-{\bf k})$ & $\mathsf{I}{\cal U}_{s}({\bf k})\mathsf{I}^{\dagger}={\cal U}_{s}(-{\bf k})$\tabularnewline
		\hline 
		$x$-Reflection & $\mathsf{M}_{x}=\sigma_{z}\otimes\sigma_{x}$ & $\begin{array}{c}
			\mathsf{M}_{x}{\cal H}_{\alpha}(k_{x},k_{y})\mathsf{M}_{x}^{\dagger}\\
			={\cal H}_{\alpha}(-k_{x},k_{y})
		\end{array}$ & $\begin{array}{c}
			\mathsf{M}_{x}{\cal U}_{s}(k_{x},k_{y})\mathsf{M}_{x}^{\dagger}\\
			={\cal U}_{s}(-k_{x},k_{y})
		\end{array}$\tabularnewline
		\hline 
		$y$-Reflection & $\mathsf{M}_{y}=\sigma_{x}\otimes\sigma_{0}$ & $\begin{array}{c}
			\mathsf{M}_{y}{\cal H}_{\alpha}(k_{x},k_{y})\mathsf{M}_{y}^{\dagger}\\
			={\cal H}_{\alpha}(k_{x},-k_{y})
		\end{array}$ & $\begin{array}{c}
			\mathsf{M}_{y}{\cal U}_{s}(k_{x},k_{y})\mathsf{M}_{y}^{\dagger}\\
			={\cal U}_{s}(k_{x},-k_{y})
		\end{array}$\tabularnewline
		\hline 
		$x$-Projective & \multirow{2}{*}{$\mathsf{L}_{x}=\sigma_{z}\otimes\begin{pmatrix}0 & 1\\
				e^{ik_{x}} & 0
			\end{pmatrix}$} & \multirow{2}{*}{$\mathsf{L}_{x}{\cal H}_{\alpha}({\bf k})\mathsf{L}_{x}^{\dagger}={\cal H}_{\alpha}({\bf k})$} & \multirow{2}{*}{$\mathsf{L}_{x}{\cal U}_{s}({\bf k})\mathsf{L}_{x}^{\dagger}={\cal U}_{s}({\bf k})$}\tabularnewline
		Translation &  &  & \tabularnewline
		\hline 
	\end{tabular}
\end{table*}

The Floquet operator ${\cal U}_{s}({\bf k})$ ($s=1,2$) in symmetric time frame
can be block diagonalized by a unitary transformation ${\cal V}(k_{x})$,
which takes the explicit form
\begin{equation}
	{\cal V}(k_{x})=\frac{1}{\sqrt{2}}\begin{bmatrix}-e^{\frac{i}{4}k_{x}} & e^{-\frac{i}{4}k_{x}} & 0 & 0\\
		0 & 0 & e^{\frac{i}{4}k_{x}} & e^{-\frac{i}{4}k_{x}}\\
		e^{\frac{i}{4}k_{x}} & e^{-\frac{i}{4}k_{x}} & 0 & 0\\
		0 & 0 & -e^{\frac{i}{4}k_{x}} & e^{-\frac{i}{4}k_{x}}
	\end{bmatrix}.\label{eq:Vkx}
\end{equation}
Specially, we have ${\cal H}'_{\alpha}({\bf k})={\cal V}(k_{x}){\cal H}_{\alpha}({\bf k}){\cal V}^{\dagger}(k_{x})$
for $\alpha=1,2,3$, where the transformed Hamiltonian components
\begin{alignat}{1}
	{\cal H}'_{0}({\bf k})= & -\sigma_{z}\otimes h_{0}(k_{x}),\label{eq:Hp1k}\\
	{\cal H}'_{1}({\bf k})= & -\sigma_{0}\otimes h_{1}(k_{y}),\label{eq:Hp2k}\\
	{\cal H}'_{2}({\bf k})= & -\sigma_{0}\otimes h_{2}(k_{y}).\label{eq:Hp3k}
\end{alignat}
We define $h_{0}(k_{x})=2J\cos(k_{x}/2)\sigma_{z}$. The
$h_{1,2}(k_{y})$ are given by the Eqs.~(\ref{eq:h1kSSH}) and (\ref{eq:h2kSSH})
with $k\rightarrow k_{y}$. The transformed Floquet operator ${\cal U}'_{s}({\bf k})={\cal V}(k_{x}){\cal U}_{s}({\bf k}){\cal V}^{\dagger}(k_{x})$
has the block-diagonal form
\begin{equation}
	{\cal U}'_{s}({\bf k})=\begin{bmatrix}{\cal U}'_{s\uparrow}({\bf k}) & 0\\
		0 & {\cal U}'_{s\downarrow}({\bf k})
	\end{bmatrix},\qquad s=1,2,\label{eq:Upsk}
\end{equation}
and the $2\times2$ components ${\cal U}'_{s\uparrow(\downarrow)}({\bf k})$
are given by
\begin{alignat}{1}
	{\cal U}'_{1\uparrow}({\bf k})&=e^{\frac{i}{2}h_{20}^{+}({\bf k})}e^{ih_{10}^{+}({\bf k})}e^{\frac{i}{2}h_{20}^{+}({\bf k})},\label{eq:U1uk}\\
	{\cal U}'_{1\downarrow}({\bf k})&=e^{\frac{i}{2}h_{20}^{-}({\bf k})}e^{ih_{10}^{-}({\bf k})}e^{\frac{i}{2}h_{20}^{-}({\bf k})},\label{eq:U1dk}\\
	{\cal U}'_{2\uparrow}({\bf k})&=e^{\frac{i}{2}h_{10}^{+}({\bf k})}e^{ih_{20}^{+}({\bf k})}e^{\frac{i}{2}h_{10}^{+}({\bf k})},\label{eq:U2uk}\\
	{\cal U}'_{2\downarrow}({\bf k})&=e^{\frac{i}{2}h_{10}^{-}({\bf k})}e^{ih_{20}^{-}({\bf k})}e^{\frac{i}{2}h_{10}^{-}({\bf k})},\label{eq:U2dk}
\end{alignat}
where
\begin{alignat}{1}
	h_{10}^{\pm}({\bf k})& \equiv h_{1}(k_{y})\pm h_{0}(k_{x}),\label{eq:h10pm}\\
	h_{20}^{\pm}({\bf k})& \equiv h_{2}(k_{y})\pm h_{0}(k_{x}).\label{eq:h20pm}
\end{alignat}
Meanwhile, the chiral symmetry and PTS operators are transformed by
${\cal V}(k_{x})$ to $\mathsf{S}'={\cal V}(k_{x})\mathsf{S}{\cal V}^{\dagger}(k_{x})$
and $\mathsf{L}'_{x}={\cal V}(k_{x})\mathsf{L}_{x}{\cal V}^{\dagger}(k_{x})$,
which are explicitly given by
\begin{alignat}{1}
	\mathsf{S}'& = -\sigma_{x}\otimes\sigma_{z},\label{eq:Sp}\\
	\mathsf{L}'_{x}& = -e^{ik_{x}/2}\sigma_{z}\otimes\sigma_{0}.\label{eq:Lxp}
\end{alignat}
We can now elaborate on the conditions for the presence of Floquet
M\"obius topological phases in our system. 

First, it can be shown that each Floquet eigenstate of the ${\cal U}'_{s}({\bf k})$
($s=1,2$) is twofold degenerate. In fact, the block-diagonal structure
of ${\cal U}'_{s}({\bf k})$ allows us to denote its four normalized
eigenstates as
\begin{alignat}{1}
	|\psi_{s\uparrow}^{\pm}({\bf k})\rangle& = [a_{s\uparrow}^{\pm}({\bf k}),b_{s\uparrow}^{\pm}({\bf k}),0,0]^{\top},\label{eq:psisu}\\
	|\psi_{s\downarrow}^{\pm}({\bf k})\rangle& = [0,0,c_{s\downarrow}^{\pm}({\bf k}),d_{s\downarrow}^{\pm}({\bf k})]^{\top}.\label{eq:psisd}
\end{alignat}
The two-component vectors $[a_{s\uparrow}^{\pm}({\bf k}),b_{s\uparrow}^{\pm}({\bf k})]^{\top}$
form two right eigenstates of the upper block ${\cal U}'_{s\uparrow}({\bf k})$
with opposite quasienergies $E_{\uparrow}^{\pm}({\bf k})=\pm E_{\uparrow}({\bf k})$,
where $E_{\uparrow}({\bf k})=\arccos\{{\rm Tr}[{\cal U}'_{s\uparrow}({\bf k})]/2\}\in[0,\pi]$.
Similarly, the vectors $[c_{s\downarrow}^{\pm}({\bf k}),d_{s\downarrow}^{\pm}({\bf k})]^{\top}$
form two right eigenstates of the lower block ${\cal U}'_{s\downarrow}({\bf k})$
with opposite quasienergies $E_{\downarrow}^{\pm}({\bf k})=\pm E_{\downarrow}({\bf k})$,
where $E_{\downarrow}({\bf k})=\arccos\{{\rm Tr}[{\cal U}'_{s\downarrow}({\bf k})]/2\}\in[0,\pi]$.
Under the operation of chiral symmetry $\mathsf{S}'$ in Eq.~(\ref{eq:Sp}), the eigenstates
$|\psi_{s\uparrow}^{\pm}({\bf k})\rangle$ are mapped to $[0,0,-a_{s\uparrow}^{\pm}({\bf k}),b_{s\uparrow}^{\pm}({\bf k})]^{\top}$
with quasienergies $\mp E_{\uparrow}({\bf k})$. As the system ${\cal U}'_{s}({\bf k})$
has only four Floquet bands under the PBC, the states $[0,0,-a_{s\uparrow}^{\pm}({\bf k}),b_{s\uparrow}^{\pm}({\bf k})]^{\top}$
must be identical to the Floquet eigenstates $|\psi_{s\downarrow}^{\mp}({\bf k})\rangle$
of ${\cal U}'_{s}({\bf k})$ up to global phase factors. It also implies
that $\pm E_{\uparrow}({\bf k})=\pm E_{\downarrow}({\bf k})=\pm E({\bf k})$
at each quasimomentum ${\bf k}=(k_{x},k_{y})$. Therefore, the eigenstates $|\psi_{s\uparrow}^{+}({\bf k})\rangle$
and $|\psi_{s\downarrow}^{+}({\bf k})\rangle$ {[}$|\psi_{s\uparrow}^{-}({\bf k})\rangle$
and $|\psi_{s\downarrow}^{-}({\bf k})\rangle${]} are degenerate at
the quasienergy $+E({\bf k})$ {[}$-E({\bf k})${]}. The twofold degeneracy
of each quasienergy level is thus proved.

Next, due to the PTS $\mathsf{L}'_{x}$, the ${\cal U}'_{s}({\bf k})$ in Eq.~(\ref{eq:Upsk})
($s=1,2$) could not goes back to itself under a $2\pi$-translation
in $k_{x}$. In fact, the Floquet operator changes as ${\cal U}'_{s}(k_{x}+2\pi,k_{y})={\cal T}_{x}{\cal U}'_{s}(k_{x},k_{y}){\cal T}_{x}^{\dagger}$,
where ${\cal T}_{x}=\sigma_{x}\otimes\sigma_{0}$. This transformation
leads to the interchange of the upper and lower blocks of ${\cal U}'_{s}({\bf k})$,
i.e., ${\cal U}'_{s\uparrow}(k_{x}+2\pi,k_{y})={\cal U}'_{s\downarrow}(k_{x},k_{y})$,
and vice versa. If the two groups of bands $E_{\uparrow}^{\pm}({\bf k})$
and $E_{\downarrow}^{\pm}({\bf k})$ have an intersection at some
quasimomentum $k_{x}$, they must be interconnected at the edges of
the first Brillouin zone $k_{x}\in[0,2\pi]$. An eigenstate initialized
in one of the upper ($+$) band will evolve smoothly to a lower band
($-$) under an ``adiabatic change'' of $k_{x}$ over $2\pi$, and
going back to itself after an additional $2\pi$ ``adiabatic change''
in $k_{x}$. This ``period-doubling'' indicates the formation of
a M\"obius twist among the two groups of bands $E_{\uparrow}^{\pm}({\bf k})$
and $E_{\downarrow}^{\pm}({\bf k})$ in momentum space. However, such
a M\"obius twist could not exist if the upper and lower groups of Floquet bands
are gapped at the quasienergies zero and $\pi$ for $k_{x}\in[0,2\pi]$.
Nevertheless, M\"obius twists between Floquet edge bands vs $k_{x}$
may still develop when the PBC and OBC are taken along the $x$ and
$y$ directions, respectively. This could happen so long as there
are high-symmetry points in $k_{x}\in[0,2\pi]$, where the upper and
lower groups of edge bands could meet and be switched. In the following,
we demonstrate that this is indeed the case due to the edge-band degeneracy
enforced by an emergent chiral symmetry at certain $k_{x}$, where
the origin of Floquet M\"obius bands can be explained by the topology
of the reduced 1D PQSSH model in Sec.~\ref{subsec:PQSSH}.

At the quasimomentum $k_{x}=\pi$, the diagonal blocks of Floquet
operator ${\cal U}'_{s}({\bf k})$ ($s=1,2$) in Eq.~(\ref{eq:Upsk}) reduces to
\begin{alignat}{1}
	{\cal U}'_{1\uparrow(\downarrow)}(\pi,k_{y})& = e^{\frac{i}{2}h_{2}(k_{y})}e^{ih_{1}(k_{y})}e^{\frac{i}{2}h_{2}(k_{y})},\label{eq:kxpi1}\\
	{\cal U}'_{2\uparrow(\downarrow)}(\pi,k_{y})& = e^{\frac{i}{2}h_{1}(k_{y})}e^{ih_{2}(k_{y})}e^{\frac{i}{2}h_{1}(k_{y})}.\label{eq:kxpi2}
\end{alignat}
All these blocks possess an emergent 1D chiral symmetry ${\cal S}=\sigma_{z}$,
in the sense that ${\cal S}{\cal U}'_{s\beta}(\pi,k_{y}){\cal S}=[{\cal U}'_{s\beta}(\pi,k_{y})]^{\dagger}$
for $s=1,2$ and $\beta=\uparrow,\downarrow$. Importantly, this chiral
symmetry is exactly the same as the chiral symmetry of the 1D PQSSH
chain in Sec.~\ref{subsec:PQSSH}. In fact, by identifying $k_{y}\leftrightarrow k$
and $J_{1,2}\leftrightarrow-J_{1,2}$, the operators in Eqs.~(\ref{eq:kxpi1})
and (\ref{eq:kxpi2}) become equivalent to the symmetrized Floquet
operators of the 1D PQSSH model in Eqs.~(\ref{eq:U1kSSH}) and (\ref{eq:U2kSSH}),
implying that they share the same topological characterizations in
terms of the winding numbers $(\omega_{0},\omega_{\pi})$. Therefore,
under the OBC along the $y$ direction, the 1D subsystem ${\cal U}'_{s\beta}(k_{x}=\pi)$
holds a pair of degenerate edge modes at the quasienergy zero $(\pi)$
if and only if $|\tan(J_{2}/2)|>|\tan(J_{1}/2)|$ ($|\tan(J_{1}/2)\tan(J_{2}/2)|>1$).
The numbers of edge modes are counted by the topological invariants
$(\omega_{0},\omega_{\pi})$ in Eq.~(\ref{eq:ome0p}) according to
the rule of bulk-edge correspondence in Eq.~(\ref{eq:BBCPQSSH}),
just as the Floquet edge modes in the 1D PQSSH model. Going back to
the 2D $\pi$-flux PQSSH model with the PBC (OBC) taken along the
$x$ ($y$) direction, there is a high-symmetry momentum $k_{x}=\pi$
where a fourfold degeneracy among Floquet edge bands at the quasienergy
zero and $\pi$ is developed as long as $|\tan(J_{2}/2)|>|\tan(J_{1}/2)|$
and $|\tan(J_{1}/2)\tan(J_{2}/2)|>1$, respectively. Through this
high-symmetry point, Floquet edge bands of the 2D $\pi$-flux PQSSH
model could form M\"obius twists in their quasienergies and then become
interchanged at $k_{x}=2\pi$. Such M\"obius twists could only be created
or destroyed by making the bulk Floquet bands to touch at the high-symmetry
momentum $k_{x}=\pi$. Moreover, this touching could appear at either
the quasienergy zero, $\pi$ or both, making it different from and
much richer than the static M\"obius topological insulators \cite{MTI07}.
Finally, the phase boundaries between different Floquet M\"obius topological
phases are still described by the Eq.~(\ref{eq:PBSSH}) of the 1D
model. The phase structure of the 2D $\pi$-flux PQSSH model is thus
formally identical to the 1D PQSSH model in Fig.~\ref{fig:PQSSH1}.
Nevertheless, the physical properties of their respective topological
phases are totally different, as will be illustrated in the following
subsections.

\subsection{Floquet M\"obius edge bands\label{subsec:FMEB}}

In this subsection, we demonstrate the existence of Floquet M\"obius
topological phases in our 2D $\pi$-flux PQSSH model through numerical
calculations. We first investigate the quasienergy spectrum of $\hat{{\cal U}}$
{[}Eq.~(\ref{eq:calU}){]} by taking the PBC and OBC along the $x$
and $y$ directions of the lattice. In this representation, the Hamiltonian
components $\hat{{\cal H}}_{0}$, $\hat{{\cal H}}_{1}$ and $\hat{{\cal H}}_{2}$
{[}Eqs.~(\ref{eq:H0})--(\ref{eq:H2}){]} are given by
\begin{alignat}{1}
	\hat{{\cal H}}_{0}(k_{x})& = J\sum_{k_{x},n}(\hat{a}_{k_{x},n}^{\dagger}\hat{b}_{k_{x},n}+e^{ik_{x}}\hat{b}_{k_{x},n}^{\dagger}\hat{a}_{k_{x},n}+{\rm H.c.})\nonumber \\
	& + J\sum_{k_{x},n}(\hat{c}_{k_{x},n}^{\dagger}\hat{d}_{k_{x},n}+e^{ik_{x}}\hat{d}_{k_{x},n}^{\dagger}\hat{c}_{k_{x},n}+{\rm H.c.}),\label{eq:H0kx}\\
	\hat{{\cal H}}_{1}(k_{x})& =J_{1}\sum_{k_{x},n}(\hat{a}_{k_{x},n}^{\dagger}\hat{c}_{k_{x},n}-\hat{b}_{k_{x},n}^{\dagger}\hat{d}_{k_{x},n}+{\rm H.c.}),\label{eq:H1kx}\\
	\hat{{\cal H}}_{2}(k_{x})& =J_{2}\sum_{k_{x},n}(\hat{c}_{k_{x},n}^{\dagger}\hat{a}_{k_{x},n+1}-\hat{d}_{k_{x},n}^{\dagger}\hat{b}_{k_{x},n+1}+{\rm H.c.}),\label{eq:H2kx}
\end{alignat}
where $k_{x}\in[0,2\pi]$ and $n=1,...,N_{y}$, with $N_{y}$ being
the total number of unit cells along $y$. The resulting Floquet operator
takes the form
\begin{equation}
	\hat{{\cal U}}(k_{x})=e^{-i\left[\hat{{\cal H}}_{0}(k_{x})+\hat{{\cal H}}_{2}(k_{x})\right]}e^{-i\left[\hat{{\cal H}}_{0}(k_{x})+\hat{{\cal H}}_{1}(k_{x})\right]}.\label{eq:calUkx}
\end{equation}
The spectrum and eigenstates of $\hat{{\cal U}}(k_{x})$ can now be
obtained by solving the eigenvalue equation $\hat{{\cal U}}(k_{x})|\psi(k_{x})\rangle=e^{-iE(k_{x})}|\psi(k_{x})\rangle$,
where the quasienergy dispersion $E(k_{x})$ is defined in the first
Floquet Brillouin zone $E\in[-\pi,\pi)$.

\begin{figure}
	\begin{centering}
		\includegraphics[scale=0.5]{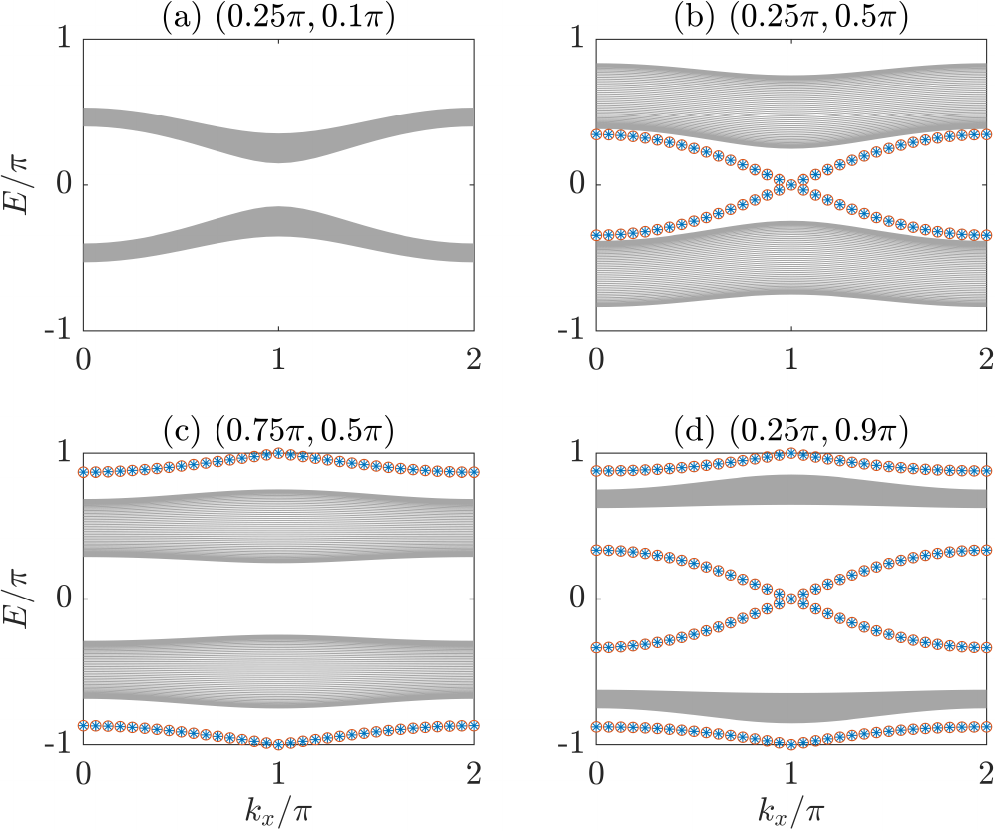}
		\par\end{centering}
	\caption{The quasienergy spectrum $E(k_{x})$ of FMTIs, obtained under the
		PBC (OBC) along the $x$ ($y$) direction and with $N_{y}=50$ unit
		cells along $y$. Gray lines, blue stars and red circles denote bulk
		states, left-localized edge states and right-localized edge states
		along the $y$ direction. The hopping amplitude along the $x$ direction
		is $J=0.1\pi$ for all panels. The hopping amplitudes $(J_{1},J_{2})$
		along the $y$ direction are given in each figure panel. (a) exemplifies
		a trivial phase, with no Floquet M\"obius edge bands. (b) displays a
		Floquet $0$-M\"obius topological insulator, with a pair of M\"obius edge
		bands twisting around the quasienergy $E=0$. (c) showcases a Floquet
		$\pi$-M\"obius topological insulator, with a pair of M\"obius edge bands
		twisting around the quasienergy $E=\pi$. (d) represents a Floquet
		$0\pi$-M\"obius topological insulator, with two pairs of M\"obius edge
		bands twisting separately around the quasienergies $E=0$ and $E=\pi$.
		\label{fig:FMTI1}}
\end{figure}
\begin{figure}
	\begin{centering}
		\includegraphics[scale=0.5]{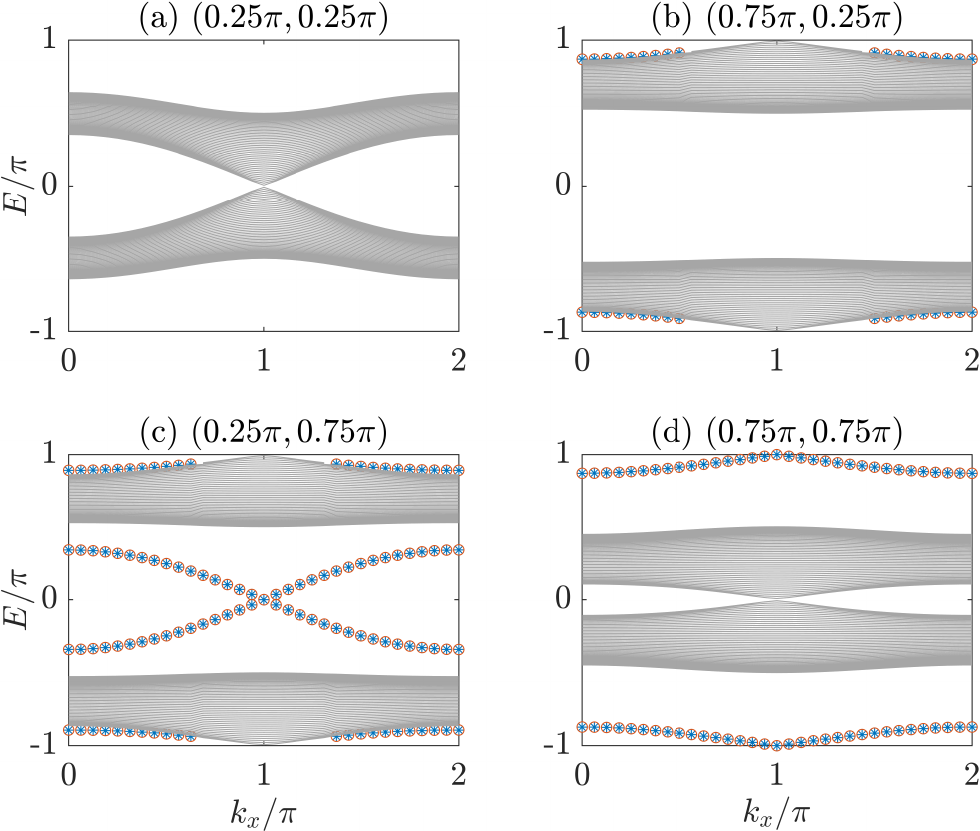}
		\par\end{centering}
	\caption{The quasienergy spectrum $E(k_{x})$ of the 2D $\pi$-flux PQSSH model
		at gapless critical points, obtained under the PBC (OBC) along the
		$x$ ($y$) direction and with $N_{y}=50$ unit cells along $y$.
		Gray lines, blue stars and red circles denote bulk states, left-localized
		edge states and right-localized edge states along the $y$ direction.
		The hopping amplitude along the $x$ direction is $J=0.1\pi$ for
		all panels. The hopping amplitudes $(J_{1},J_{2})$ along the $y$
		direction are given in the caption of each figure panel. (a) and (b)
		represent trivial critical points, with Floquet bulk bands touching
		at $E=0$ and $E=\pi$, respectively. (c) displays a topological critical
		point, with bulk Floquet bands touching at $E=\pi$ and M\"obius edge
		bands retaining around $E=0$. (d) shows another topological critical
		point, with bulk Floquet bands touching at $E=0$ and M\"obius edge
		bands interconnecting around $E=\pi$. \label{fig:FMTI2}}
\end{figure}

In Fig.~\ref{fig:FMTI1}, we present typical quasienergy spectra of
the 2D $\pi$-flux PQSSH model in its four distinct insulating phases.
The hopping amplitude $J=0.1\pi$ along $x$ and the number of unit
cells $N_{y}=50$ along $y$ directions are used throughout. The spectrum
in Fig.~\ref{fig:FMTI1}(a) has no signatures of M\"obius edge bands.
The system parameters $(J_{1},J_{2})$ are taken in the bottom region
of the phase diagram Fig.~\ref{fig:PQSSH1}, where the topological
invariants $(\omega_{0},\omega_{\pi})=(0,0)$. Therefore, this phase
region indeed corresponds to a trivial Floquet insulator. The spectrum
in Fig.~\ref{fig:FMTI1}(b) has a pair of M\"obius edge bands crossing
at $k_{x}=\pi$ with a fourfold degeneracy point at the quasienergy
zero. The system parameters $(J_{1},J_{2})$ are taken in the left
region of the phase diagram Fig.~\ref{fig:PQSSH1}, where the topological
invariants $(\omega_{0},\omega_{\pi})=(1,0)$. This phase corresponds
to a Floquet topological insulator with M\"obius edge bands twisting
along $k_{x}$ at the quasienergy zero. We thus refer to it as Floquet
$0$-M\"obius topological insulator. The spectrum in Fig.~\ref{fig:FMTI1}(c)
has a pair of M\"obius edge bands crossing at $k_{x}=\pi$ with a fourfold
degeneracy at the quasienergy $\pi$. The system parameters $(J_{1},J_{2})$
are taken in the right region of Fig.~\ref{fig:PQSSH1}, where the
topological invariants $(\omega_{0},\omega_{\pi})=(0,1)$. This phase
describes a Floquet topological insulator with M\"obius edge bands twisting
along $k_{x}$ at the quasienergy $\pi$. We thus refer to it as Floquet
$\pi$-M\"obius topological insulator. Finally, the spectrum in 
Fig.~\ref{fig:FMTI1}(d) has two pairs of M\"obius edge bands crossing at
$k_{x}=\pi$ with two fourfold degeneracy points at both the quasienergies
zero and $\pi$. The system parameters $(J_{1},J_{2})$ are taken
in the top region of Fig.~\ref{fig:PQSSH1}, where the topological
invariants $(\omega_{0},\omega_{\pi})=(1,1)$. This phase describes
a Floquet topological insulator with M\"obius edge bands twisting along
$k_{x}$ at both the quasienergies zero and $\pi$. We refer to it
as Floquet $0\pi$-M\"obius topological insulator. Importantly, even
though the trivial and $0$-M\"obius topological insulators can exist
in static systems \cite{MTI07}, the $\pi$-M\"obius and $0\pi$-M\"obius
topological insulators have M\"obius twisted edge bands at the quasienergy
$\pi$, which are not available in non-driven cases. Therefore, Floquet
time-periodic drivings could indeed generate unique types of M\"obius
topological insulators beyond equilibrium.

In Fig.~\ref{fig:FMTI2}, we present typical quasienergy spectra of
the 2D $\pi$-flux PQSSH model along its critical lines $J_{2}=J_{1}$
and $J_{2}=\pi-J_{1}$. The hopping amplitude $J=0.1\pi$ along $x$
and the number of unit cells $N_{y}=50$ along $y$ directions are
used throughout. The spectrum in Fig.~\ref{fig:FMTI2}(a) has no M\"obius
edge bands. The system parameters are taken along $J_{2}=J_{1}$ with
$J_{1}\in(0,\pi/2)$ in Fig.~\ref{fig:PQSSH1}, where the bulk gap
closes at the quasienergy zero and the topological invariants $(\omega_{0},\omega_{\pi})=(0,0)$.
It thus describes a trivial critical point. The spectrum in Fig.~\ref{fig:FMTI2}(b)
also has no M\"obius edge bands. The system parameters are taken along
$J_{2}=\pi-J_{1}$ with $J_{1}\in[\pi/2,\pi)$ in Fig.~\ref{fig:PQSSH1},
where the bulk gap closes at the quasienergy $\pi$ and the topological
invariants $(\omega_{0},\omega_{\pi})=(0,0)$. It thus describes another
type of trivial critical point, which is yet unique to Floquet systems
due to the gaplessness of bulk bands at $E=\pm\pi$. The spectrum
in Fig.~\ref{fig:FMTI2}(c) shows a pair of interconnected M\"obius
edge bands twisting at $k_{x}=\pi$ with a fourfold degeneracy at
the quasienergy zero. The system parameters are taken along $J_{2}=\pi-J_{1}$
with $J_{1}\in(0,\pi/2)$ in Fig.~\ref{fig:PQSSH1}, where the bulk
gap closes at the quasienergy $E(k_{x}=\pi)=\pm\pi$ and the topological
invariants $(\omega_{0},\omega_{\pi})=(1,0)$. It thus describes a
topologically nontrivial critical point with coexisting gapless bulk
bands and Floquet M\"obius edge bands around the quasienergy zero. We
refer to these edge states as $0$-M\"obius critical edge bands. Although
these edge bands could appear in the static counterpart of our model,
they could not persist at its critical point \cite{MTI07}. Therefore,
the critical line $J_{2}=\pi-J_{1}$ with $J_{1}\in(0,\pi/2)$ constitutes
Floquet-induced topological critical points with $0$-M\"obius edge
bands. 
Finally, the spectrum in Fig.~\ref{fig:FMTI2}(d) shows a pair
of M\"obius edge bands crossing at $k_{x}=\pi$ with a fourfold degeneracy
at the quasienergy $\pi$. The system parameters are taken along $J_{2}=J_{1}$
with $J_{1}\in(\pi/2,\pi)$ in Fig.~\ref{fig:PQSSH1}, where the bulk
gap closes at the quasienergy $E(k_{x}=\pi)=0$ and the topological
invariants $(\omega_{0},\omega_{\pi})=(0,1)$. It thus describes a
topologically nontrivial critical point with coexisting gapless bulk
bands at quasienergy zero and Floquet M\"obius edge bands around the quasienergy $\pi$.
We refer to the latter edge states as $\pi$-M\"obius critical edge bands,
which are anomalous and unique to driven systems. In the sense of
gapless Floquet topology \cite{FgSPT01}, they offer defining signatures
of Floquet-enriched topological critical points that are protected
by chiral symmetry and $\mathbb{Z}_2$-PTS in 2D systems.

\begin{table*}
	\caption{Topological properties of the 2D $\pi$-flux PQSSH model. $J_{1}$
		and $J_{2}$ are strengths of intracell and intercell hopping amplitudes
		along the $y$ direction. The topological invariants $(\omega_{0},\omega_{\pi})$
		are defined in Eq.~(\ref{eq:ome0p}). The numbers of Floquet M\"obius
		edge bands $(N_{0},N_{\pi})$ around zero and $\pi$ quasienergies
		are related to the invariants $(\omega_{0},\omega_{\pi})$ according
		to the rule of bulk-edge correspondence in Eq.~(\ref{eq:BBCPQSSH}).
		\label{tab:2}}
	
	\centering{}%
	\begin{tabular}{c|c|c|c}
		\hline 
		\multirow{2}{*}{\textbf{Phases}} & \multirow{2}{*}{\textbf{Conditions}} & \textbf{Topological} & \textbf{Floquet M\"obius}\tabularnewline
		&  & \textbf{invariants} & \textbf{edge bands}\tabularnewline
		\hline 
		\hline 
		\multirow{2}{*}{Trivial} & $|\tan(J_{2}/2)|\leq|\tan(J_{1}/2)|$ & \multirow{2}{*}{$(\omega_{0},\omega_{\pi})=(0,0)$} & \multirow{2}{*}{$(N_{0},N_{\pi})=(0,0)$}\tabularnewline
		& $|\tan(J_{1}/2)\tan(J_{2}/2)|\leq1$ &  & \tabularnewline
		\hline 
		Floquet $0$-M\"obius & $|\tan(J_{2}/2)|>|\tan(J_{1}/2)|$ & \multirow{2}{*}{$(\omega_{0},\omega_{\pi})=(1,0)$} & \multirow{2}{*}{$(N_{0},N_{\pi})=(2,0)$}\tabularnewline
		topological & $|\tan(J_{1}/2)\tan(J_{2}/2)|\leq1$ &  & \tabularnewline
		\hline 
		Floquet $\pi$-M\"obius & $|\tan(J_{2}/2)|\leq|\tan(J_{1}/2)|$ & \multirow{2}{*}{$(\omega_{0},\omega_{\pi})=(0,1)$} & \multirow{2}{*}{$(N_{0},N_{\pi})=(0,2)$}\tabularnewline
		topological & $|\tan(J_{1}/2)\tan(J_{2}/2)|>1$ &  & \tabularnewline
		\hline 
		Floquet $0\pi$-M\"obius & $|\tan(J_{2}/2)|>|\tan(J_{1}/2)|$ & \multirow{2}{*}{$(\omega_{0},\omega_{\pi})=(1,1)$} & \multirow{2}{*}{$(N_{0},N_{\pi})=(2,2)$}\tabularnewline
		topological & $|\tan(J_{1}/2)\tan(J_{2}/2)|>1$ &  & \tabularnewline
		\hline 
	\end{tabular}
\end{table*}

In Table \ref{tab:2}, we summarize the main topological properties
of the 2D $\pi$-flux PQSSH model. The hopping amplitude along the
$x$ direction is assumed to be $J\neq0$ in order to define a 2D
system. The trivial phase in Table \ref{tab:2} includes the gapless
critical lines $J_{2}=J_{1}$ with $J_{1}\in(0,\pi/2)$, $J_{2}=\pi-J_{1}$
with $J_{1}\in[\pi/2,\pi)$, and the gapped bulk phase with $(\omega_{0},\omega_{\pi})=(0,0)$
in Fig.~\ref{fig:PQSSH1}. In all these phases, there are no Floquet
M\"obius edge bands ($N_{0}=N_{\pi}=0$) crossing the quasienergies
zero and $\pi$. The Floquet $0$-M\"obius topological phase in the
second row of Table \ref{tab:2} includes the gapless critical line
$J_{2}=\pi-J_{1}$ with $J_{1}\in(0,\pi/2)$ and the gapped bulk phase
with $(\omega_{0},\omega_{\pi})=(1,0)$ in Fig.~\ref{fig:PQSSH1}.
In these phases, there is a pair of M\"obius edge bands crossing the
quasienergy zero ($N_{0}=2$) at $k_{x}=\pi$. The Floquet $\pi$-M\"obius
topological phase in the third row of Table \ref{tab:2} includes
the gapless critical line $J_{2}=J_{1}$ with $J_{1}\in(\pi/2,\pi)$
and the gapped bulk phase with $(\omega_{0},\omega_{\pi})=(0,1)$
in Fig.~\ref{fig:PQSSH1}. In these phases, there is a pair of M\"obius
edge bands crossing the quasienergy $\pi$ ($N_{\pi}=2$) at $k_{x}=\pi$.
The Floquet $0\pi$-M\"obius topological phase in the last row of Table
\ref{tab:2} includes the gapped insulator phase with $(\omega_{0},\omega_{\pi})=(1,1)$
in Fig.~\ref{fig:PQSSH1}, in which there are two pairs of M\"obius
edge bands crossing the quasienergies zero and $\pi$ ($N_{0}=N_{\pi}=2$)
at $k_{x}=\pi$, separately. It deserves to mention that despite FMTIs,
we also identify two topologically nontrivial critical lines with
coexisting gapless bulk spectra and interconnected M\"obius edge bands.
Signified by the presence of these critical M\"obius zero and $\pi$
edge bands, these gapless phase boundaries form extensions of the
recently discovered Floquet topological critical points \cite{FgSPT01}
to 2D driven systems with chiral and projective translational symmetries.
The M\"obius edge bands surrounding the quasienergy $\pi$ are also
unique to Floquet systems, regardless of whether the bulk spectrum
is gapped or gapless.

Up to now, we have verified the presence of gapped and gapless Floquet
M\"obius topological phases with zero/$\pi$ M\"obius edge bands in the
2D $\pi$-flux PQSSH model. Their bulk topological invariants and
bulk-edge correspondence can be applied to characterize M\"obius topological
phases in other 2D Floquet systems with chiral and projective translational
symmetries. In the next subsection, we provide further evidence for
M\"obius topological phases and M\"obius edge bands in our Floquet system from
the viewpoints of entanglement spectra and adiabatic dynamics.

\subsection{Entanglement spectrum and dynamics\label{subsec:EAD}}

\begin{figure}
	\begin{centering}
		\includegraphics[scale=0.49]{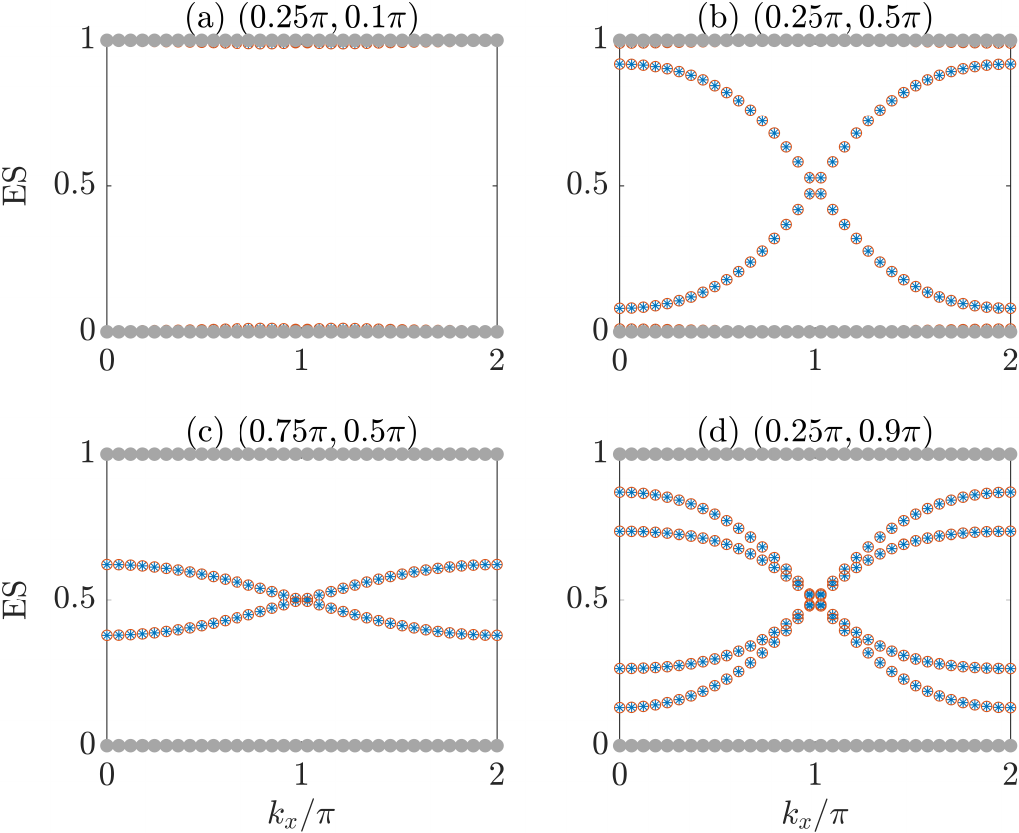}
		\par\end{centering}
	\caption{The entanglement spectrum of gapped Floquet M\"obius phases, with the
		PBC taken along both the $x$, $y$ directions and $N_{y}=50$ unit
		cells along $y$. An equal bipartition is taken in real space along
		the $y$ direction. The system is prepared at half-filling in each
		case. The hopping amplitude along $x$ is $J=0.1\pi$ for all panels.
		The hopping amplitudes $(J_{1},J_{2})$ along $y$ are given in the
		caption of each panel. Gray dots, blue stars and red circles denote
		extended, left-localized and right-localized eigenmodes along the
		$y$ direction of the Floquet entanglement Hamiltonian $H_{{\rm X}}$.
		\label{fig:FMTI3}}
\end{figure}

\begin{figure}
	\begin{centering}
		\includegraphics[scale=0.5]{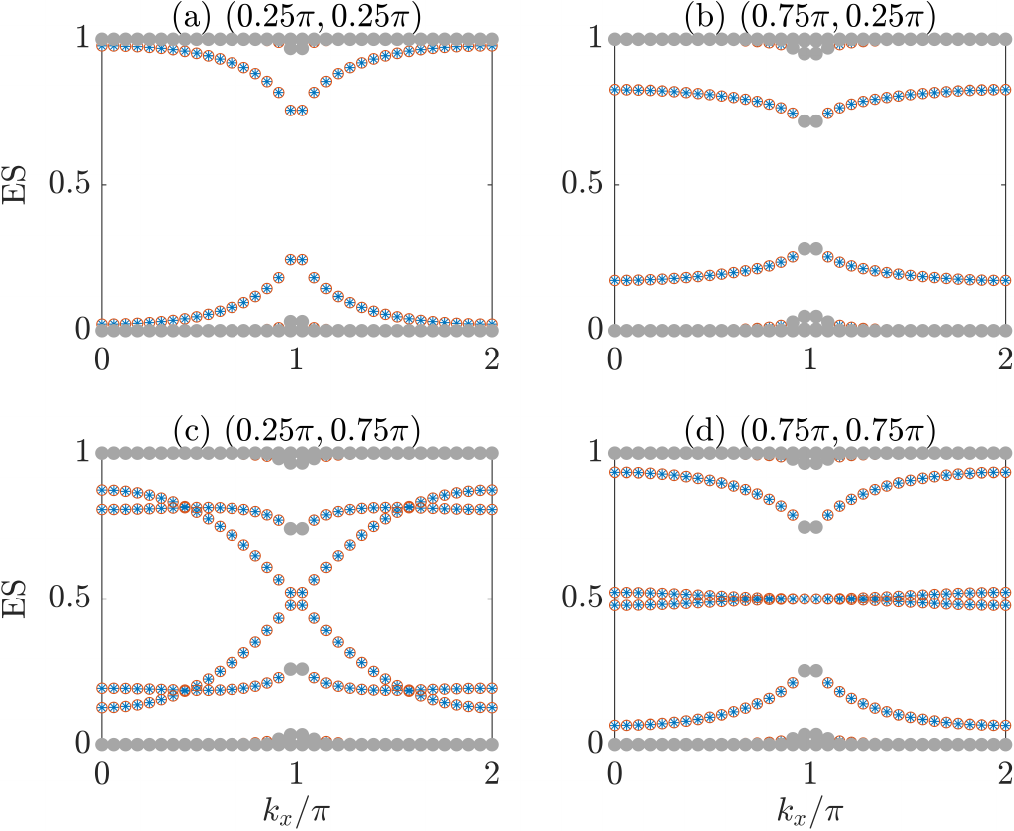}
		\par\end{centering}
	\caption{The entanglement spectrum at the gapless critical points between different
		Floquet M\"obius insulators, with the PBC taken along both the $x$,
		$y$ directions and $N_{y}=50$ unit cells along $y$. An equal bipartition
		is taken in real space along the $y$ direction, and the system is
		prepared at half-filling in each case. The hopping amplitude along
		$x$ is $J=0.1\pi$ for all panels. The hopping amplitudes $(J_{1},J_{2})$
		along $y$ are given in the caption of each panel. Gray dots, blue
		stars and red circles denote extended, left-localized and right-localized
		eigenmodes along the $y$ direction of the Floquet entanglement Hamiltonian
		$H_{{\rm X}}$. \label{fig:FMTI4}}
\end{figure}

The entanglement spectrum (ES) of a Floquet-driven system can be obtained
following the recipe of Refs.~\cite{FgSPT01,ESEE01}. For our system,
let us denote $\{|\psi_{\lambda}(k_{x})\rangle\}$ as the set of occupied
(occ.) eigenstates of the Floquet operator $\hat{{\cal U}}(k_{x})$
in Eq.~(\ref{eq:calUkx}). At half-filling, the normalized Floquet
many-particle state of the system can be expressed as $|\Psi\rangle=\prod_{\lambda\in{\rm occ.}}|\psi_{\lambda}(k_{x})\rangle\equiv\prod_{\lambda\in{\rm occ.}}\hat{\psi}_{\lambda}^{\dagger}(k_{x})|\emptyset\rangle$,
where $|\emptyset\rangle$ denotes the vacuum state. The corresponding
density matrix reads $\rho=|\Psi\rangle\langle\Psi|$. Under the PBC
along both $x$ and $y$ directions, we decompose our system into
two equal parts X and Y. As we work in the momentum space along $x$,
this bipartition corresponds to introducing spatial entanglement cuts
along $y$ at the unit cells $n=1,N_{y}/2$. Tracing out all the degrees
of freedom belonging to the subsystem Y (with unit cell indices $n=N_{y}/2+1,...,N_{y}$
along $y$), we obtain the reduced density matrix of subsystem X (with
unit cell indices $n=1,...,N_{y}/2$ along $y$) as $\rho_{{\rm X}}\equiv{\rm Tr}_{{\rm Y}}\rho=e^{-H_{{\rm X}}}/{\rm Tr}e^{-H_{{\rm X}}}$.
Here, $H_{{\rm X}}$ is the Floquet entanglement Hamiltonian, whose
eigenspectrum forms the ES of our system at half-filling and under
equal bipartition. As the density matrix $\rho$ represents a Gaussian
state, the ES can be numerically obtained from the eigenspectrum of
the single-particle correlation matrix \cite{ESEE01}.

In Figs.~\ref{fig:FMTI3}(a)--\ref{fig:FMTI3}(d), we show the typical
ES obtained in the symmetric time frame for the gapped phases in our
2D $\pi$-flux PQSSH model. The respective physical parameters are
in one-to-one correspondence with those of Figs.~\ref{fig:FMTI1}(a)--\ref{fig:FMTI1}(d).
The ES is presented via the correlation matrix spectrum in each case,
which is in one-to-one correspondence with the ES \cite{ESEE01}.
In Fig.~\ref{fig:FMTI3}(a), we find no signatures of M\"obius edge
bands in the ES, verifying that this case corresponds to a trivial Floquet insulator.
In both Figs.~\ref{fig:FMTI3}(b) and \ref{fig:FMTI3}(c), we find
one pair of M\"obius edge bands in the ES. They are localized at the
left and right entanglement cuts and twisted at the high-symmetry
quasimomentum $k_{x}=\pi$. They offer entanglement signatures for
the Floquet $0$-M\"obius and $\pi$-M\"obius edge bands in Figs.~\ref{fig:FMTI1}(b)
and \ref{fig:FMTI1}(c), respectively. Note in passing that as the
ES is restricted in the range of $[0,1]$, one cannot directly determine
whether a M\"obius edge band in the ES is originated from a $0$-M\"obius
or a $\pi$-M\"obius Floquet topological insulator. Finally, we observe
two pairs of M\"obius edge bands in the ES of Fig.~\ref{fig:FMTI3}(d),
which are all crossed at the high-symmetry momentum $k_{x}=\pi$.
They provide entanglement signatures for the coexistence of Floquet
$0$-M\"obius and $\pi$-M\"obius quasienergy edge bands in Figs.~\ref{fig:FMTI1}(d),
offering further evidence for the presence of Floquet $0\pi$-M\"obius
topological insulators in our system.

In Fig.~\ref{fig:FMTI4}, we show the typical ES at four gapless critical
points in our 2D $\pi$-flux PQSSH model, obtained also in the symmetric
time frame. The system parameters used in Figs.~\ref{fig:FMTI4}(a)--\ref{fig:FMTI4}(d)
are in one-to-one correspondence with those of Figs.~\ref{fig:FMTI2}(a)--\ref{fig:FMTI2}(d).
In Figs.~\ref{fig:FMTI4}(a) and \ref{fig:FMTI4}(b), we find no M\"obius
edge bands crossing the center of ES. As mentioned before, these two
cases correspond to topologically trivial critical points without
M\"obius edge bands. Their trivial topology are thus confirmed by the
ES results. In Figs.~\ref{fig:FMTI4}(c) and \ref{fig:FMTI4}(d),
we find one pair of M\"obius edge bands crossing the center of ES at
the high-symmetry momentum $k_{x}=\pi$ in each case. They provide
entanglement signals for the $0$-M\"obius and $\pi$-M\"obius critical
Floquet edge bands in Figs.~\ref{fig:FMTI2}(c) and \ref{fig:FMTI2}(d),
respectively. Putting together, we conclude that both the gapped and
gapless Floquet M\"obius topological phases of our 2D $\pi$-flux PQSSH
model could leave signatures in the ES via the presence of M\"obius
twisted entanglement edge bands. The crossing point of these edge
bands at $k_{x}=\pi$ in the ES is protected by the emergent subsystem
chiral symmetry of the Floquet operator at this high-symmetry momentum,
and further characterized by the winding numbers $(\omega_{0},\omega_{\pi})$
of the reduced 1D PQSSH model defined there.

In Sec.~\ref{subsec:BMT}, we have mentioned that one may realize
adiabatic exchange between different M\"obius edge bands at the same
edge without utilizing bulk states. We now demonstrate this explicitly
with numerical calculations. We consider the dynamical process in
which the quasimomentum $k_{x}$ works as an adiabatic parameter along
a synthetic dimension, and it is changed only stroboscopically. Due
to the M\"obius twist, a full adiabatic cycle contains two periods in
$k_{x}$, i.e., the change of $k_{x}$ from $0$ to $4\pi$. We divide
this period into $M$ steps and define $k_{x}(j)=4\pi j/M$ for $j=0,1,...,M-1,M$.
Starting at $k_{x}(0)=0$, the stroboscopic time-evolution over such
an adiabatic cycle is determined by the propagator 
\begin{equation}
	{\hat F}_{M}=\mathbb{T}\left[\prod_{j=1}^{M}\hat{{\cal U}}(k_{x}(j))\right],\label{eq:FM}
\end{equation}
where $\mathbb{T}$ performs time-ordering and each $\hat{{\cal U}}(k_{x}(j))$
is obtained by substituting $k_{x}(j)=4\pi j/M$ into the Eq.~(\ref{eq:calUkx}).
The adiabatic limit is reached by taking $M\rightarrow\infty$.

\begin{figure}
	\begin{centering}
		\includegraphics[scale=0.51]{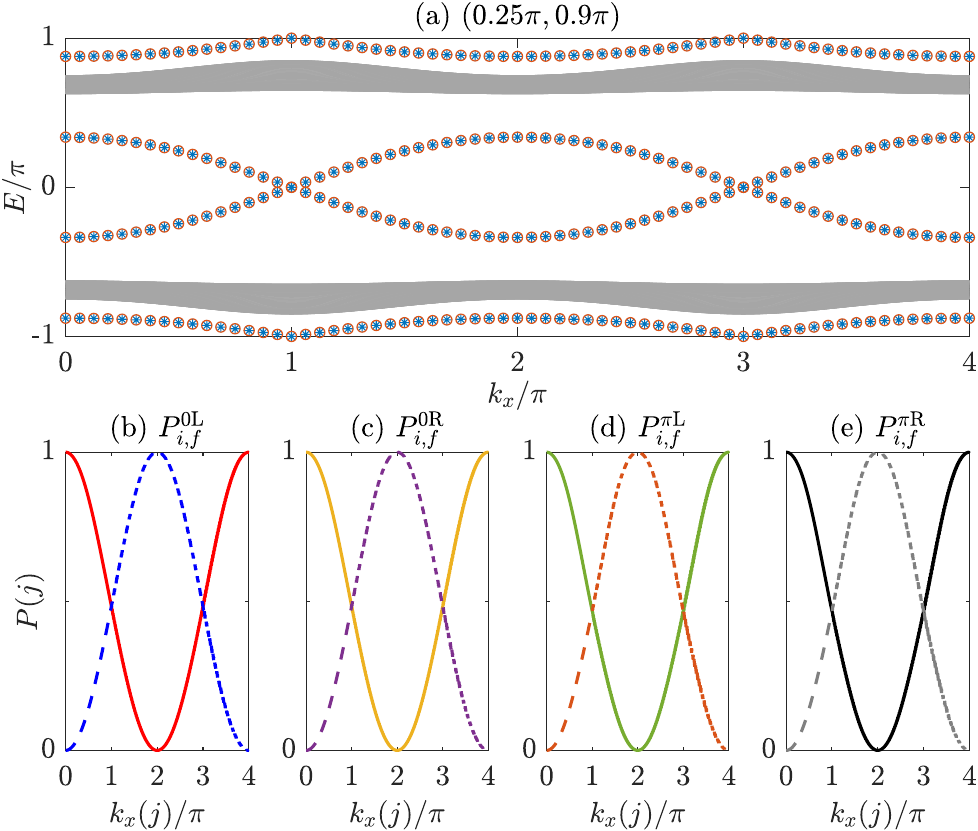}
		\par\end{centering}
	\caption{Adiabatic dynamics of Floquet M\"obius edge states. The PBC (OBC) is
		taken along the $x$ ($y$) direction, and there are $N_{y}=50$ unit
		cells along $y$. System parameters are $J=0.1\pi$, $J_{1}=0.25\pi$
		and $J_{2}=0.9\pi$ for all panels. (a) shows the Floquet spectrum
		of $\hat{{\cal U}}(k_{x})$ over two periods in $k_{x}$. Gray lines,
		blue stars and red circles denote bulk extended states, left-localized
		edge states and right-localized edge states along the $y$ direction.
		In (b)--(e), $k_{x}(j)=4\pi j/M$ is the adiabatic parameter and
		$j=0,1,...,M-1,M$ are indices of adiabatic evolution steps. The number
		of steps in each adiabatic cycle $k_{x}(j):0\rightarrow4\pi$ is set
		to $M=4\times10^{5}$. $P(j)$ gives the population of the evolved
		state in the initial state {[}solid lines in (b)--(e){]} and in its
		initial M\"obius partner state {[}dashed lines in (b)--(e){]} at each step
		$j$. In (b), the initial state and its M\"obius partner state are the left
		edge modes in the $0$-gap below and above $E=0$ at $k_{x}(0)=0$,
		respectively. In (c), the initial state and its M\"obius partner state are
		the right edge modes in the $0$-gap below and above $E=0$ at $k_{x}(0)=0$,
		respectively. In (d), the initial state and its M\"obius partner state are
		the left edge modes in the $\pi$-gap above $E=-\pi$ and below $E=\pi$
		at $k_{x}(0)=0$, respectively. In (e), the initial state and its
		M\"obius partner state are the right edge modes in the $\pi$-gap above $E=-\pi$
		and below $E=\pi$ at $k_{x}(0)=0$. \label{fig:FMTI5}}
\end{figure}

Initially, the system is prepared in a Floquet M\"obius eigenstate of
$\hat{{\cal U}}(k_{x}(0))$ with quasienergy $-E$ at the left or
right edge of the system. Due to the PTS, this initial state has a
M\"obius partner with quasienergy $+E$, which is localized at the same
edge. If this edge-state pair resides in the gap around the quasienergy
$0$ ($\pi$) and localizes at the left edge, we denote them as $|\psi_{i}^{0{\rm L}}\rangle$
and $|\psi_{f}^{0{\rm L}}\rangle$ ($|\psi_{i}^{\pi{\rm L}}\rangle$
and $|\psi_{f}^{\pi{\rm L}}\rangle$), where ``L'' means left, ``$i$''
means initial and ``$f$'' means final. If this edge-state pair
resides in the gap around the quasienergy $0$ ($\pi$) and localizes
at the right edge, we denote them as $|\psi_{i}^{0{\rm R}}\rangle$
and $|\psi_{f}^{0{\rm R}}\rangle$ ($|\psi_{i}^{\pi{\rm R}}\rangle$
and $|\psi_{f}^{\pi{\rm R}}\rangle$), where ``R'' means right.
After the evolution over $j$ adiabatic steps in which $k_{x}$ goes
from $0$ to $4\pi j/M$, we obtain the populations of the evolved
state in the initial state $P_{i}^{\epsilon{\rm Z}}(j)$ and in its
initial M\"obius partner state $P_{f}^{\epsilon{\rm Z}}(j)$ as
\begin{alignat}{1}
	P_{i}^{\epsilon{\rm Z}}(j)&=|\langle\psi_{i}^{\epsilon{\rm Z}}|{\hat F}_{j}|\psi_{i}^{\epsilon{\rm Z}}\rangle|^{2},\label{eq:Pi}\\
	P_{f}^{\epsilon{\rm Z}}(j)&=|\langle\psi_{f}^{\epsilon{\rm Z}}|{\hat F}_{j}|\psi_{i}^{\epsilon{\rm Z}}\rangle|^{2},\label{eq:Pf}
\end{alignat}
where $\epsilon=0,\pi$, ${\rm Z=L,R}$, and 
\begin{equation}
	{\hat F}_{j}\equiv\hat{{\cal U}}(k_{x}(j))\hat{{\cal U}}(k_{x}(j-1))\cdots\hat{{\cal U}}(k_{x}(2))\hat{{\cal U}}(k_{x}(1)).\label{eq:Fj}
\end{equation}

If there is a M\"obius twist between a pair of edge bands at $k_{x}=\pi$
through the quasienergy $\epsilon$, we would expect the population
$P_{i}^{\epsilon{\rm Z}}(j)$ ($P_{f}^{\epsilon{\rm Z}}(j)$) to decrease
(increase) from $1$ ($0$) to $1/2$ with the increase of $j$ from
$0$ to $M/4$ during the stroboscopic dynamics. Crossing the twist
at $k_{x}(M/4)=\pi$, the $P_{i}^{\epsilon{\rm Z}}(j)$ ($P_{f}^{\epsilon{\rm Z}}(j)$)
will further decrease (increase) to $0$ ($1$) with the increase
of $j$ from $M/4+1$ to $M/2$, where $k_{x}(M/2)=2\pi$. Therefore,
after one period in $k_{x}$, the populations of the initial state
in itself and in its initial M\"obius partner state are interchanged,
so that the initial state is pumped to its orthogonal M\"obius partner
state at the same edge. It also means that the two Floquet M\"obius
edge bands are switched after one period in $k_{x}$. Such a process
is impossible in the absence of the M\"obius twist at $k_{x}=\pi$.
This process will be reversed with the further increase of $j$ from
$M/2+1$ to $M$, so that $k_{x}(j)$ goes from $2\pi$ to $4\pi$
to accomplish a second period. After reaching $k_{x}(M)=4\pi$, we
obtain $P_{i}^{\epsilon{\rm Z}}(M)=1=P_{i}^{\epsilon{\rm Z}}(0)$
and $P_{f}^{\epsilon{\rm Z}}(M)=0=P_{f}^{\epsilon{\rm Z}}(0)$ as
the initial conditions. Therefore, although the Floquet operator of
the system is $2\pi$-periodic in $k_{x}$, the $P_{i}^{\epsilon{\rm Z}}$
and $P_{f}^{\epsilon{\rm Z}}$ are both $4\pi$-periodic in $k_{x}$.
This ``periodic doubling'' in dynamics, as manifested in the adiabatic evolution
of edge-band populations $P_{i}^{\epsilon{\rm Z}}$ and $P_{f}^{\epsilon{\rm Z}}$,
offers dynamical signatures for the presence of M\"obius edge
bands.

To be explicit, we consider a numerical example in which the system
holds Floquet M\"obius edge bands surrounding both the quasienergies
zero and $\pi$. The Floquet spectrum of $\hat{{\cal U}}(k_{x})$
in this case is shown in Fig.~\ref{fig:FMTI5}(a). It is obtained
under the PBC and OBC along $x$ and $y$ directions. Two periods
of the spectrum in $k_{x}$ are presented to illustrate a complete
cycle of the M\"obius edge bands. The adiabatic dynamics of edge-state
populations are shown in Figs.~\ref{fig:FMTI5}(b)--\ref{fig:FMTI5}(e).
The solid (dashed) lines show the $P_{i}^{0{\rm L}}(j)$, $P_{i}^{0{\rm R}}(j)$,
$P_{i}^{\pi{\rm L}}(j)$ and $P_{i}^{\pi{\rm R}}(j)$ ($P_{f}^{0{\rm L}}(j)$,
$P_{f}^{0{\rm R}}(j)$, $P_{f}^{\pi{\rm L}}(j)$ and $P_{f}^{\pi{\rm R}}(j)$)
versus the slowly varied $k_{x}(j)$ in each corresponding figure
panel. We find the periodic doubling of edge-band dynamics in all
cases, verifying our theoretical expectations. The observed adiabatic
dynamics confirms that there are indeed two pairs of Floquet $0$-M\"obius
and $\pi$-M\"obius edge bands coexisting in the case considered in
Fig.~\ref{fig:FMTI5}(a). Floquet M\"obius edge bands in other cases
of the 2D $\pi$-flux PQSSH model can be dynamically characterized
in the same way. In experiments, we may also use the periodic doubling
of edge-state dynamics to demonstrate the presence of Floquet M\"obius
edge bands in relevant setups like acoustic waveguides~\cite{MTI10},
in which the populations of Floquet edge states can be directly imaged
over different driving periods.

\section{Conclusion and discussion\label{sec:Sum}}
In this work, we uncovered and characterized unique M\"obius topological
phases in Floquet-driven systems. Focusing on a 2D periodically quenched
SSH model with a $\pi$ magnetic flux per plaquette, we revealed the
presence of interconnected M\"obius edge bands twisting around the quasienergy
zero, $\pi$, or both at a high-symmetry point in the Floquet Brillouin
zone. These M\"obius phases are protected by the combined chiral and
projective translational symmetries, going beyond the standard tenfold
way of topological insulators. They are further characterized by a
pair of generalized winding numbers $(\omega_{0},\omega_{\pi})$ defined
at the high-symmetry momentum $k_{x}=\pi$, where an emergent 1D chiral
symmetry enforces fourfold degeneracies among Floquet edge states
at zero/$\pi$ quasienergies with M\"obius twists. Under the OBC, the
invariants $(\omega_{0},\omega_{\pi})$ count the numbers of M\"obius
edge bands $(N_{0},N_{\pi})$ around zero and $\pi$ quasienergies
via the bulk-edge correspondence $(N_{0},N_{\pi})=2(|\omega_{0}|,|\omega_{\pi}|)$.
We also unveiled topologically nontrivial critical points where M\"obius
edge bands around zero/$\pi$ quasienergies can coexist with gapless
bulk spectra. These $0$-M\"obius and $\pi$-M\"obius critical edge states
can be characterized on an equal footing with the FMTIs. Finally, we demonstrated
our findings through numerical investigations of the Floquet spectrum,
entanglement spectrum and adiabatic edge-band dynamics, yielding results
consistent with the theoretical predictions.

In the absence of the $\pi$ magnetic flux, the static and Floquet systems in Fig.~\ref{fig:Sketch}(b) and Figs.~\ref{fig:Sketch}(c)--\ref{fig:Sketch}(d) reduce to the 2D extension of the SSH model and its Floquet counterpart, respectively. The topological properties of these 2D models have been studied in Ref.~\cite{FloCry04}. In short, the 2D $0$-flux SSH model possesses two distinct phases at half-filling. One of them is a trivial insulator, whereas the other represents a second-order topological insulator (SOTI) with four degenerate zero modes localized at the four corners of the lattice. The SOTI phase is reached if $|J_2|>|J_1|$ and $|J'_x|>|J_x|$. The 2D $0$-flux PQSSH model holds four distinct phases at half-filling, with three of them representing Floquet SOTIs. When $|J'_x|<|J_x|$, the system is topologically trivial with no Floquet corner modes. When $|J'_x|>|J_x|$, the system holds four Floquet corner modes at quasienergy zero if $|\tan(J_2/2)|>|\tan(J_1/2)|$ and $|\tan(J_1/2)\tan(J_2/2)|<1$, four Floquet corner modes at quasienergy $\pi$ if $|\tan(J_2/2)|<|\tan(J_1/2)|$ and $|\tan(J_1/2)\tan(J_2/2)|>1$, and four Floquet corner modes at both quasienergies zero and $\pi$ if $|\tan(J_2/2)|>|\tan(J_1/2)|$ and $|\tan(J_1/2)\tan(J_2/2)|>1$. Notably, there are no M\"obius edge bands in all these static and Floquet second-order topological phases. The $\pi$-flux, which introduces the $\mathbb{Z}_2$-PTS is thus essential to generate M\"obius topological phases.

Our discovery extends the study of M\"obius topological phases to nonequilibrium
driven systems and enriches the symmetry classification of Floquet
topological states. Moreover, the $\pi$-M\"obius and $0\pi$-M\"obius
topological insulators are associated with phase transitions at topologically
nontrivial critical points of Floquet origin, making them unique to
periodically driven systems. In experiments, the 2D Floquet model
we considered is within reach of current quantum or quantum-inspired
simulators. For example, in acoustic waveguides, the time dimension
can be simulated by an extra spatial dimension along $z$-direction
in the paraxial wave equation of sound pressure. The time-periodically
quenched hopping amplitudes can be simulated by adding link tubes
between intracell and intercell waveguides alternatively along the
propagation direction $z$ (i.e., the synthetic time dimension) of sound waves. Furthermore, the M\"obius
edge bands might be detected via direct phase measurements of the
Floquet operator and the imaging of interference population dynamics
of the edge states. Therefore, the setup realized in Refs.~\cite{MTI10,MTI11}
may serve as a good starting point to experimentally explore our found
Floquet M\"obius topological phases. In practice, the adiabatic switching
dynamics of Floquet M\"obius edge bands allows the evolution of one
topological edge state to another at the same edge through the symmetry-protected
M\"obius twist. One may thus utilize two Floquet M\"obius edge modes at the same edge to form
the two levels of a robust local qubit that can be adiabatically controlled, which may
find applications in certain quantum information and computation tasks.

\begin{acknowledgments}
	This work is supported by the National Natural Science Foundation of China (Grants No.~12275260 and No.~11905211), the Fundamental Research Funds for the Central Universities (Grant No.~202364008), and the Young Talents Project of Ocean University of China.
\end{acknowledgments}
\vspace{0.5cm}

\appendix

\section{Topology of the SSH model}\label{sec:SSH}

The SSH model describes noninteracting fermions in a 1D tight-binding
lattice with staggered nearest-neighbor hoppings \cite{SSH}. It has
two sublattices A and B in each unit cell. The Hamiltonian of SSH
model takes the form
\begin{equation}	\hat{H}=\sum_{j}\left(J_{1}\hat{a}_{j}^{\dagger}\hat{b}_{j}+J_{2}\hat{b}_{j}^{\dagger}\hat{a}_{j+1}+{\rm H.c.}\right),\label{eq:HSSH}
\end{equation}
where $j\in\mathbb{Z}$ is the unit cell index. $\hat{a}_{j}^{\dagger}$
($\hat{b}_{j}^{\dagger}$) creates a fermion in the sublattice A (B).
$J_{1}$ and $J_{2}$ are intracell and intercell hopping amplitudes.
When the intracell coupling is stronger ($|J_{2}|<|J_{1}|$), the
system is topologically trivial, with no degenerate edge modes at
zero energy  under the open boundary condition (OBC). When the intercell coupling is stronger ($|J_{2}|>|J_{1}|$),
the system is topologically nontrivial, with a pair of degenerate
edge modes at zero energy under the OBC. A topological phase transition happens
when $|J_{2}|=|J_{1}|$, where the system is also topologically trivial
with no edge-localized zero modes under the OBC.

\begin{figure}
	\begin{centering}
		\includegraphics[scale=0.47]{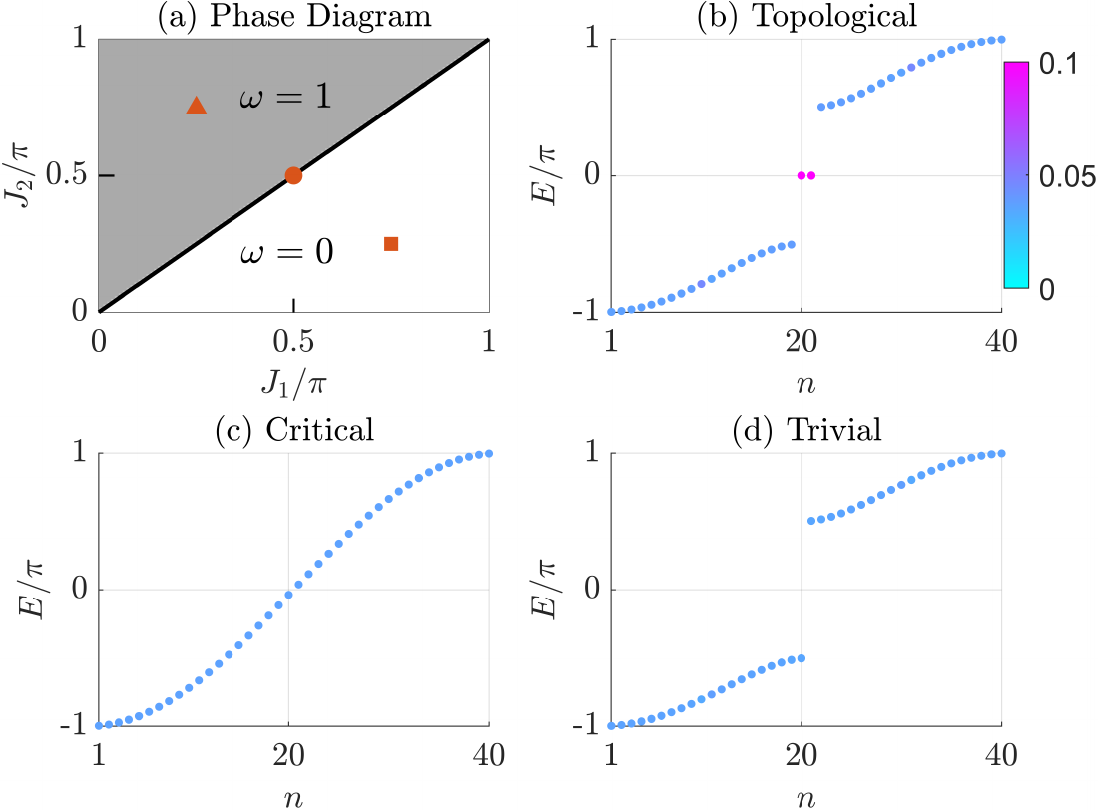}
		\par\end{centering}
	\caption{Phase diagram {[}in (a){]} and OBC spectra {[}in (b)--(d){]} of the
		SSH model. The winding number $\omega$ for the topological
		and trivial insulator phases are shown explicitly in (a). (b) Spectrum
		in the topological phase, with $(J_{1},J_{2})=(0.25\pi,0.75\pi)$
		as denoted by the $\blacktriangle$ in (a). (c) Spectrum at the critical
		point, with $(J_{1},J_{2})=(0.5\pi,0.5\pi)$ as denoted by the $\CIRCLE$
		in (a). (d) Spectrum in the trivial phase, with $(J_{1},J_{2})=(0.75\pi,0.25\pi)$
		as denoted by the $\blacksquare$ in (a). In (b)--(d), we take $20$
		unit cells in the calculation of the OBC spectra $E$.
		$n$ is the state index. The color of each data point is given by
		the inverse participation ratio of the related state.\label{fig:SSH}}
\end{figure}

To distinguish different topological phases of the SSH model, one may take the
periodic boundary condition (PBC) and transform $\hat{H}$ from the
position to momentum representations through $\hat{a}_{j}=\frac{1}{\sqrt{N}}\sum_{k}e^{ikj}\hat{a}_{k}$
and $\hat{b}_{j}=\frac{1}{\sqrt{N}}\sum_{k}e^{ikj}\hat{b}_{k}$. Here,
$N$ denotes the total number of unit cells. $k\in[-\pi,\pi)$ is
the quasimomentum. In $k$-space, the Hamiltonian can be expressed
as $\hat{H}=\sum_{k}\hat{\Psi}_{k}^{\dagger}H(k)\hat{\Psi}_{k}$,
where $\hat{\Psi}_{k}^{\dagger}\equiv(\hat{a}_{k}^{\dagger},\hat{b}_{k}^{\dagger})$
and the Bloch Hamiltonian
\begin{alignat}{1}
	H(k) & =\begin{pmatrix}0 & J_{1}+J_{2}e^{-ik}\\
		J_{1}+J_{2}e^{ik} & 0
	\end{pmatrix}\label{eq:HkSSH}\\
	& =(J_{1}+J_{2}\cos k)\sigma_{x}+J_{2}\sin k\sigma_{y}.\nonumber 
\end{alignat}
We use $\sigma_{0}$ to denote the $2\times2$ identity matrix and
$\sigma_{x,y,z}$ to denote three Pauli matrices. The bulk spectrum
of $H(k)$ is given by $E_{\pm}(k)=\pm\sqrt{(J_{1}+J_{2}\cos k)^{2}+(J_{2}\sin k)^{2}}$.
It is clear $H(k)$ has the chiral symmetry ${\cal S}=\sigma_{z}$
with ${\cal S}H(k){\cal S}=-H(k)$ and ${\cal S}^{2}=\sigma_{0}$,
the time-reversal symmetry ${\cal T}=\sigma_{0}$ with ${\cal T}H^{*}(k){\cal T}^{\dagger}=H(-k)$
and ${\cal T}^{2}=\sigma_{0}$, the particle-hole symmetry ${\cal C}=\sigma_{z}$
with ${\cal C}H^{*}(k){\cal C}^{\dagger}=-H(-k)$ and ${\cal C}^{2}=\sigma_{0}$,
and the inversion symmetry ${\cal I}=\sigma_{x}$ with ${\cal I}H(k){\cal I}^{\dagger}=H(-k)$.
The SSH model then belongs to the symmetry class BDI \cite{TenfoldNJP},
whose topological insulating phases are characterized by an integer quantized
winding number
\begin{equation}
	w\equiv\int_{-\pi}^{\pi}\frac{dk}{2\pi}\partial_{k}\phi(k),\label{eq:wSSH}
\end{equation}
where $\phi(k)=\arctan[h_{y}(k)/h_{x}(k)]$, $h_{x}(k)=J_{1}+J_{2}\cos k$
and $h_{y}(k)=J_{2}\sin k$. Geometrically, $w$ counts the number
of times that the vector $[h_{x}(k),h_{y}(k)]$ circulates the origin
when $k$ goes over the first Brillouin zone. In the trivial and topological
gapped phases, we have
\begin{equation}
	w=\begin{cases}
		0, & |J_{2}|<|J_{1}|\\
		1, & |J_{2}|>|J_{1}|
	\end{cases}.\label{eq:wSSH2}
\end{equation}
A quantized change of $w$ happens at the transition point $|J_{2}|=|J_{1}|$,
where the trajectory of $[h_{x}(k),h_{y}(k)]$ passes through the
origin with $h_{x}(k)=h_{y}(k)=0$. To find a unified topological
characterization of both the transition point and its surrounding gapped
phases, we may extend the lower off-diagonal element of $H(k)$ to
the complex plane with $J_{1}+J_{2}e^{ik}\rightarrow f(z)=J_{1}+J_{2}z$,
where $z\in\mathbb{C}$. Counting the difference between the zeros
and poles of $f(z)$ (including their multiplicity and order) inside the unit circle gives us a topological
integer $\omega$ according to the Cauchy's argument principle \cite{gSPT0},
whose values are given by
\begin{equation}
	\omega=\begin{cases}
		0, & |J_{2}|\leq|J_{1}|\\
		1, & |J_{2}|>|J_{1}|
	\end{cases}.\label{eq:ome}
\end{equation}
We find that $\omega=w$ in the gapped phases and $\omega=0$ along the
phase boundary. One can thus utilize $\omega$ to characterize the
bulk topology of all the insulating phases and gapless critical points
of the SSH model. The bulk phase diagram of the SSH model is shown
in Fig.~\ref{fig:SSH}(a).

Signatures of bulk topology at the edges can be identified from the
spectrum and eigenstates of the SSH model under the OBC. Solving the eigenvalue equation $\hat{H}|\psi\rangle=E|\psi\rangle$
numerically with the $\hat{H}$ given by Eq.~(\ref{eq:HSSH}), we
obtain the OBC spectrum of the SSH model, with three typical cases
shown in Figs.~\ref{fig:SSH}(b)--\ref{fig:SSH}(d). We notice that
degenerate edge modes at zero energy can only appear in the topological
insulator phase with $|J_{2}|>|J_{1}|$ and $\omega=w=1$. The
localization of these zero modes can also be deduced analytically
from their wave functions in the thermodynamic limit. For example,
considering a half-infinite chain with the OBC taken at its left edge,
we get the zero energy solution of $\hat{H}|\psi\rangle=E|\psi\rangle$
as \cite{FgSPT03}
\begin{equation}
	|\psi_{{\rm L}}^{(0)}\rangle=\sum_{j=1}^{\infty}\left(-J_{1}/J_{2}\right)^{j-1}\hat{a}_{j}^{\dagger}|\emptyset\rangle,\label{eq:0SSH}
\end{equation}
where ``L'' denotes ``left'' and $|\emptyset\rangle$ denotes
the vacuum state. There is another zero mode $|\psi_{{\rm R}}^{(0)}\rangle=\sum_{j=-\infty}^{N}\left(-J_{1}/J_{2}\right)^{N-j}\hat{b}_{j}^{\dagger}|\emptyset\rangle$
occupying only the B sublattices if the OBC is also taken at the right
end, where $N\gg1$ and R denotes ``right''. These two orthogonal
modes {[}$\langle\psi_{{\rm L}}^{(0)}|\psi_{{\rm R}}^{(0)}\rangle=0${]}
form a pair of degenerate edge zero modes if and only if $|J_{2}|>|J_{1}|$.
Referring to the Eq.~(\ref{eq:ome}), we arrive at the bulk-edge correspondence
of the SSH model, i.e., $N_{0}=2|\omega|$, where $N_{0}$ is the
number of topological edge modes at zero energy in each insulating
phase or along the phase boundary.

In summary, due to the competition between intracell and intercell
hopping strengths, the SSH model could realize either a trivial band
insulator or a topological insulator phase. These two phases are separated
by a topological transition at $|J_{2}|=|J_{1}|$, i.e., when the
two hopping strengths are equal. In Sec.~\ref{subsec:PQSSH} of the
main text, we see that both the insulator phases and the critical
points of the SSH model could be greatly enriched when a Floquet driving
is applied. Moreover, phases and transitions unique to nonequilibrium
systems could emerge thanks to the driving fields.

\section{Symmetry analysis}\label{sec:symmetry}
As mentioned in the main text, the PTS and the chiral symmetry are crucial for both the M\"obius topological insulators and the FMTIs. In this Appendix, we provide a detailed discussion of these symmetries. In Appendix~\ref{App B: PTS}, we prove that a Hamiltonian with PTS can be block diagonalized, and we present the explicit construction of the corresponding unitary matrix. In Appendix~\ref{App B: chiral}, we provide a detailed proof that the Floquet operator satisfies chiral symmetry at $k_{x}=\pi$.

\subsection{Projective translational symmetry}\label{App B: PTS}
Our 2D $\pi$-flux model has the PTS, which means that its Hamiltonian satisfies the communication relation $[\mathsf{L}_{x},{\mathcal{H}}]=0$, where $\mathsf{L}_{x}$ is the PTS operator. The $\mathsf{L}_{x}$ can be diagonalized and expressed as $\mathsf{L}_{x}=VDV^{\dagger}$, where
\begin{align}
    D&=\begin{bmatrix}
        -e^{\frac{i}{2} k_{x}} & 0 & 0 & 0 \\ 
        0 & -e^{\frac{i}{2} k_{x}} & 0 & 0 \\ 
        0 & 0 & e^{\frac{i}{2} k_{x}} & 0 \\ 
        0 & 0 & 0 & e^{\frac{i}{2} k_{x}}
    \end{bmatrix},\label{App B: D}\\
    V&=\frac{1}{\sqrt{2}}\begin{bmatrix}
        -e^{-\frac{i}{4} k_{x}} & 0 & e^{-\frac{i}{4} k_{x}} & 0 \\ 
        e^{\frac{i}{4} k_{x}} & 0 & e^{\frac{i}{4} k_{x}} & 0 \\ 
        0 & e^{-\frac{i}{4} k_{x}} & 0 & -e^{-\frac{i}{4} k_{x}} \\ 
        0 & e^{\frac{i}{4} k_{x}} & 0 & e^{\frac{i}{4} k_{x}}
    \end{bmatrix}.\label{App B: V}
\end{align}
Next, substituting $\mathsf{L}_{x}=VDV^{\dagger}$ into the commutation relation, we obtain
\begin{equation}\label{App B:commu}
    [VDV^{\dagger},{\mathcal{H}}]=[D,V^{\dagger}{\mathcal{H}}V]\equiv[D,{\mathcal{H}}']=0.
\end{equation}
As seen in Eq.~(\ref{App B: D}), $\mathsf{L}_{x}$ has two pairs of doubly degenerate eigenvalues. When substituted into the constraint $D\mathcal{H}'=\mathcal{H}'D$, this immediately implies that $\mathcal{H}'$ is block-diagonal.
Therefore, the unitary matrix ${\cal V}(k_{x})$ given in Eq.~(\ref{eq:Vkx}) can be used to block-diagonalize the Hamiltonian $\mathcal{H}$, resulting in $\mathcal{H}'={\cal V}(k_{x}){\cal H}{\cal V}^{\dagger}(k_{x})$, where
\begin{equation}
\begin{split}
    {\cal V}(k_{x})&=V^{\dagger}\\
    &=\frac{1}{\sqrt{2}}
    \begin{bmatrix}-e^{\frac{i}{4}k_{x}} & e^{-\frac{i}{4}k_{x}} & 0 & 0\\
		0 & 0 & e^{\frac{i}{4}k_{x}} & e^{-\frac{i}{4}k_{x}}\\
		e^{\frac{i}{4}k_{x}} & e^{-\frac{i}{4}k_{x}} & 0 & 0\\
		0 & 0 & -e^{\frac{i}{4}k_{x}} & e^{-\frac{i}{4}k_{x}}
	\end{bmatrix}.
    \end{split}
\end{equation}
Since the Hamiltonian ${\cal H}$ consists of the terms ${\cal H}_{1}+{\cal H}_{2}+{\cal H}_{3}$, it can be directly verified that the transformation ${\cal V}(k_{x})$ can block diagonalize each ${\cal H}_{\alpha}$ individually.

\subsection{Chiral symmetry}\label{App B: chiral}
The chiral symmetry ensures the winding number in Eq.~(\ref{eq:w12}) to be a valid topological invariant for FMTIs. Here, we offer more details to prove the chiral symmetry of $\mathcal{U}'_{s\beta}$ ($s=1,2$ and $\beta=\uparrow,\downarrow$) in Eqs.~(\ref{eq:kxpi1}) and (\ref{eq:kxpi2}).

In Eqs.~(\ref{eq:U1uk})--(\ref{eq:U2dk}) we have given the expression for ${\cal U}'_{s\beta}$. As presented in the main text, $h_{0}(k_{x})=2J\cos(k_{x}/2)\sigma_{z}$, which vanishes at $k_{x}=\pi$. Therefore, we obtain ${\cal U}'_{s \uparrow}(\pi,k_{y})={\cal U}'_{s \downarrow}(\pi,k_{y})$, as shown in Eqs.~(\ref{eq:kxpi1}) and (\ref{eq:kxpi2}). It suffices to prove that ${\cal S}{\cal U}'_{s \uparrow}(\pi,k_{y}){\cal S}={\cal U}'^{\dagger}_{s \uparrow}(\pi,k_{y})$, where ${\cal S}=\sigma_{z}$. By applying Euler's formula $e^{i\theta \vec{n} \cdot \vec{\sigma}_{n}}=\cos{\theta}\sigma_{0}+i\sin{\theta}\vec{n} \cdot \vec{\sigma}_{n}$, we can expand ${\cal U}'_{1 \uparrow}(\pi,k_{y})$ and ${\cal U}'_{2 \uparrow}(\pi,k_{y})$ as
\begin{align}
    {\cal U}'_{1 \uparrow}(\pi,k_{y})={\cal J}_{10}\sigma_{0}+i\left({\cal J}_{1x} \sigma_{x}+ {\cal J}_{1y}\sigma_{y}\right),\label{equ:u1up expand}\\
    {\cal U}'_{2 \uparrow}(\pi,k_{y})={\cal J}_{20}\sigma_{0}+i\left({\cal J}_{2x} \sigma_{x}+ {\cal J}_{2y}\sigma_{y}\right).\label{equ:u2up expand}
\end{align}
where ${\cal J}_{10}={\cal J}_{20}=\cos{J_1}\cos{J_{2}}-\sin{J_{1}}\sin{J_{2}}\cos{k_{y}}$, and the other components are
\begin{equation}
    \begin{split}
    {\cal J}_{1x}=&
        \cos{J_{1}}\sin{J_{2}}\cos{k_{y}}+\sin{J_{1}}\cos^{2}{\frac{J_{2}}{2}}\\
        &-\sin{J_{1}}\sin^{2}{\frac{J_{2}}{2}}\cos{2k_{y}},\\
    {\cal J}_{1y}=&\cos{J_{1}}\sin{J_{2}}\sin{k_{y}}-\sin{J_{1}}\sin^{2}{\frac{J_{2}}{2}}\sin{2k_{y}},\\
    {\cal J}_{2x}=&\sin{J_{1}}\cos{J_{2}}+\cos{J_{1}\sin{J_{2}}\cos{k_{y}}},\\
    {\cal J}_{2y}=&\sin{J_{2}\sin{k_{y}}}.
    \end{split}
\end{equation}

For Eqs.~(\ref{equ:u1up expand}) and~(\ref{equ:u2up expand}), by using the commutation relation of Pauli matrices, we find the effect of chiral symmetry ${\cal S}=\sigma_z$ on Floquet operator ${\cal U}'_{s \uparrow}(\pi,k_{y})$ as
\begin{equation}
\begin{split}
     {\cal S}{\cal U}'_{s \uparrow}(\pi,k_{y}){\cal S}&={\cal J}_{s0}\sigma_{0}-i\left[{\cal J}_{sx} \sigma_{x}+ {\cal J}_{sy}\sigma_{y}\right]\\
     &={\cal U}'^{\dagger}_{s \uparrow}(\pi,k_{y}).
\end{split}
\end{equation}
It turns out to be equal to the Hermitian conjugate of the original Floquet operator. At $k_{x}=\pi$, the chiral symmetry of our model is strictly preserved, and the Floquet M\"obius edge bands exhibit a fourfold degeneracy at zero or $\pi$ quasienergy. When $k_{x}$ deviates from $\pi$, the M\"obius edge bands split from zero and evolve into two doubly degenerate branches. In other words, the existence of fourfold degeneracy at $k_{x}=\pi$ necessarily guarantees the presence of a M\"obius twist in edge bands. Therefore, the winding number defined at $k_{x}=\pi$ is sufficient to characterize the M\"obius topology of our Floquet system.

\section{Experimental proposal}\label{sec:exp}
As mentioned in the main text, static M\"obius topological insulators have been realized in acoustic systems~\cite{MTI10,MTI11}. The simulation of Floquet systems in acoustics is also well-established \cite{FloCry22,FloCry23}. These studies indicate that realizing FMTIs in acoustic systems is feasible. Based on the design principles in Refs.~\cite{MTI10,MTI11,FloCry22,FloCry23}, we here present an experimental proposal for our model.

\begin{figure}
    \begin{centering}
    \includegraphics[scale=0.24]{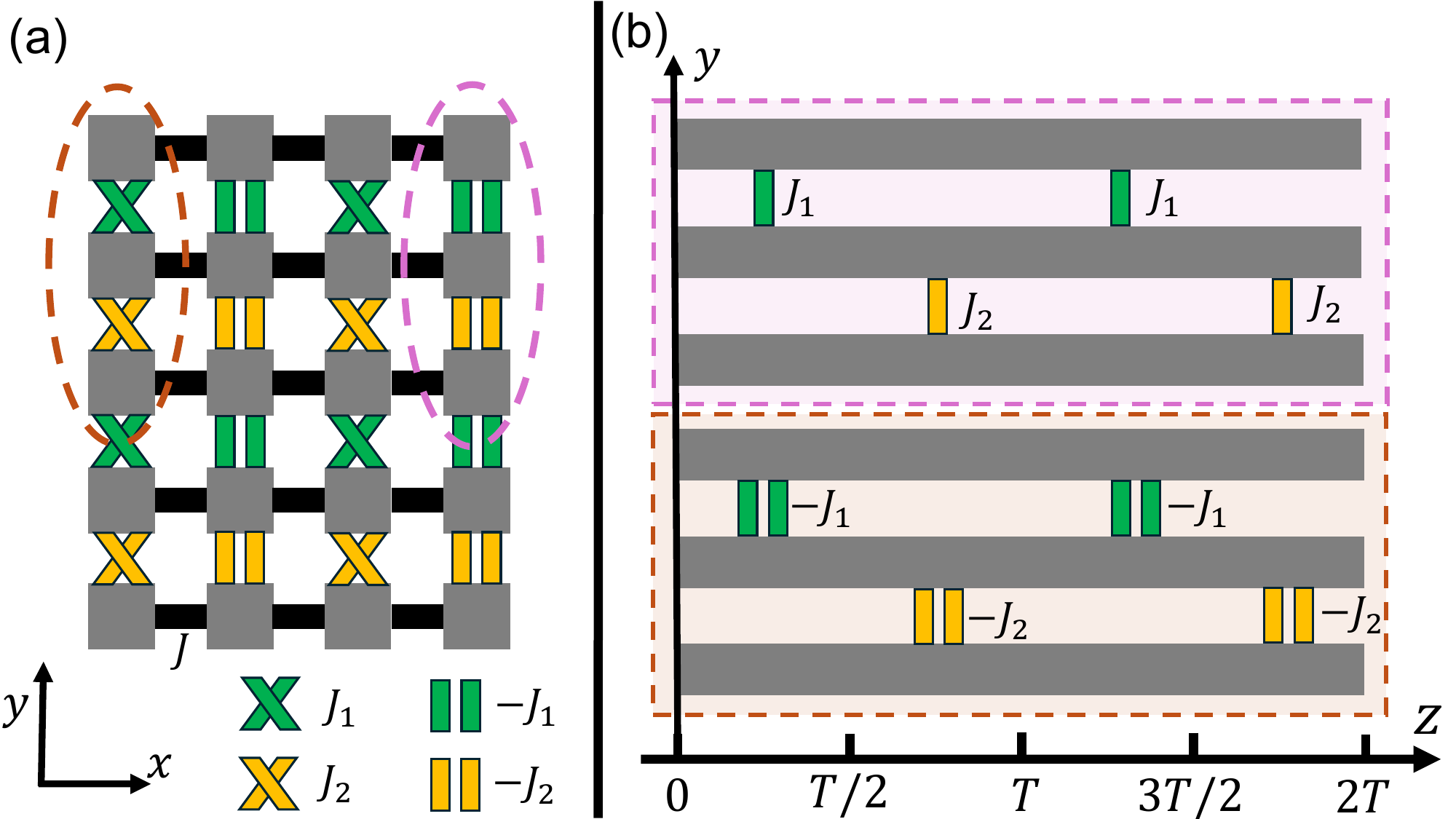}
    \par\end{centering}
    \caption{Schematic illustration of the experiment. (a) shows schematic cross-sectional view of the experimental setup in the $x-y$ plane. The gray squares denote the acoustic waveguides, while the couplings between them are implemented by narrow tubes of different sizes, corresponding to different hopping amplitudes. The black narrow tubes indicate couplings along the $x$ direction, all with strength $J$. The green and yellow tubes represent couplings along the $y$ direction with strengths $J_{1}$ and $J_{2}$, respectively. These correspond to two different coupling configurations: the crossed coupling and the parallel coupling, associated with positive and negative coupling strengths, respectively. (b) shows the $y-z$ cross-sectional views of the regions enclosed by the pink and brown dashed lines in panel (a), which are located at the upper and lower parts, respectively. Here, the $z$ axis represents the time dimension, and the coupling structure is periodically modulated along the $z$ axis to simulate a Floquet system.}
    \label{FMTI_exp_pro}
\end{figure}

The static 2D lattice model is shown in Fig.~\ref{fig:Sketch}(b). By applying the Peierls substitution, we obtain a flux-free Hamiltonian, as given by Eq.~(\ref{eq:HBBH}). In practice, the experiment does not directly realize a $\pi$-flux model. Instead, it simulates the Hamiltonian of lattice model without flux. Here, the $\pi$ flux contributes a phase factor $e^{i\pi m}$, where $m$ is the lattice index. As a result, the hopping amplitudes along the $y$ direction take opposite signs in neighboring columns, as illustrated by $\pm J_{1(2)}$ in Figs.~\ref{fig:Sketch}(b)--(d). Therefore, a key aspect of the experiment is to realize couplings with both positive and negative amplitudes. This issue has already been addressed in Ref.~\cite{PMCOUPLE}, where such a scheme was further employed to simulate a Hermitian and non-Hermitian M\"obius topological insulator~\cite{MTI11,MTI20}. As shown in Fig.~\ref{FMTI_exp_pro}(a), there exist two types of couplings along the $y$ direction: the straight-linked and the cross-linked. First, we consider the coupling between two resonators, which can be described by a $2\times2$ Hamiltonian as
\begin{equation}
    H_{c}=\begin{pmatrix}
        \omega_{0} & K\\
        K & \omega_{0}
    \end{pmatrix},
\end{equation}
where $\omega_{0}$ is the frequency of the two resonators and $K\in \mathbb{R}$ is the coupling strength. The eigenvalues are $\omega_{\pm}=\omega_{0}\pm K$, corresponding to the eigenvectors $\varphi_{\pm}=(1/\sqrt{2})(1,\pm 1)^{T}$. The signs $+$ and $-$ correspond to the in-phase and out-of-phase coupling modes, respectively. Therefore, when the eigenfrequency of the in-phase mode is higher than that of the out-of-phase mode, i.e., $\omega_{+}>\omega_{-}$, the coupling strength satisfies $K>0$. Conversely, when the eigenfrequency of the in-phase mode is lower than that of the out-of-phase mode, i.e., $\omega_{+}<\omega_{-}$, the coupling strength satisfies $K<0$. In this way, both positive and negative coupling strengths can be realized~\cite{PMCOUPLE}. Experimentally, by identifying the two resonance frequencies of the coupled cavities and determining, at the lower frequency, whether the mode corresponds to the in-phase or out-of-phase configuration, the sign of the coupling strength can be obtained. In practice, the cross-linked geometry corresponds to positive coupling, whereas the straight-linked geometry corresponds to negative coupling~\cite{PMCOUPLE}. In this way, by employing different coupling geometries, one can simulate a static M\"obius topological insulator. In the following, we extend this approach to the acoustic Floquet system.

In acoustic systems, the time dimension can be simulated by introducing an additional spatial direction into the 2D experimental platform. The paraxial wave equation can be written as
\begin{equation}
    i\frac{\partial p}{\partial z}=\left(-\frac{\nabla_{\bot}^{2}}{2k_{z}}-\frac{k^{2}-k_{z}^{2}}{2k_{z}}\right)p. \label{equ:S-like equ}
\end{equation}
where $\nabla^{2}_{\perp}\equiv \partial^{2}_{x}+\partial^{2}_{y}$, and $p$ is the acoustic pressure. By treating $z$ as the time $t$, the wave function of the system can be obtained by solving the Schr\"odinger-like equation in Eq.~(\ref{equ:S-like equ})~\cite{FloCry22,FloCry23}. Therefore, in the experiment, by employing different coupling structures at different positions along the waveguide direction ($z$ axis) to connect adjacent waveguides, one can realize the desired periodically quenched system.

The specific experimental scheme is illustrated in Fig.~\ref{FMTI_exp_pro}, where we show how to construct an acoustic waveguide array to realize our model. Fig.~\ref{FMTI_exp_pro}(a) shows schematic cross-sectional view of the experimental setup in the $x-y$ plane. By employing cross-linked and straight-linked connections, we obtain $\pm J_{1(2)}$. The green and orange thin tubes are used to distinguish the coupling strengths $J_{1}$ and $J_{2}$, respectively. The coupling along the $x$ direction remains constant and is represented by the black thin tubes. Fig.~\ref{FMTI_exp_pro}(b) shows the $y-z$ cross-sectional views of the regions enclosed by the pink and brown dashed lines in Fig.~\ref{FMTI_exp_pro}(a), each containing three waveguides. Along the $z$ direction, we take different coupling structures within different regions. In the interval $[l-1)T,(l-\tfrac{1}{2})T]$, where $l\in \mathbb{Z}^{+}$, only the coupling tubes with strength $J_{1}$ (green) are present. The pink and orange colors indicate the columns with positive and negative couplings, respectively. From this perspective, the straight-linked tubes overlap with each other, while the two cross-linked tubes appear adjacent. Similarly, In the interval $[(l-\frac{1}{2})T,lT]$, only the coupling tubes with strength $J_{2}$ (orange) are present.

Through the above scheme, the Floquet M\"obius topological phases in our system can be realized, and the distribution of their eigenstates can be obtained by measuring the acoustic pressure field.
In practice, the quenching rate can be tuned by varying the spatial distance between linking tubes along the $z$ direction. A few driving periods should not be challenging to realize in current experiments \cite{FloCry22,FloCry23}. The presence of loss can usually be expressed by a uniform background term in the system Hamiltonian, which could contribute a constant imaginary part to the Floquet spectrum. Since the topological features of the system are not affected by such a global loss, we may just ignore it and focus on the real part of Floquet spectrum in experimental detection.


\begin{thebibliography}{99}

\bibitem{FTPRev01} J. Cayssol, B. D\'ora, F. Simon,
and R. Moessner, Floquet topological insulators, Phys. Status Solidi
RRL \textbf{7}, 101 (2013).

\bibitem{FTPRev02} M. Bukov, L. D'Alessio, and A. Polkovnikov, Universal
high-frequency behavior of periodically driven systems: from dynamical
stabilization to Floquet engineering, Adv. Phys. \textbf{64}, 139
(2015).

\bibitem{FTPRev03} A. Eckardt, Colloquium: Atomic quantum gases in
periodically driven optical lattices, Rev. Mod. Phys. \textbf{89},
011004 (2017).

\bibitem{FTPRev04} T. Oka and S. Kitamura, Floquet engineering of
quantum materials, Annu. Rev. Condens. Matter Phys. \textbf{10}, 387
(2019).

\bibitem{FTPRev05} M. S. Rudner and N. H. Lindner, Band structure
engineering and non-equilibrium dynamics in Floquet topological insulators,
Nat. Rev. Phys. \textbf{2}, 229 (2020).

\bibitem{FTPRev06} F. Harper, R. Roy, M. S. Rudner, and S. L. Sondhi,
Topology and broken symmetry in Floquet systems, Annu. Rev. Condens.
Matter Phys. \textbf{11}, 345 (2020).

\bibitem{FTPRev07} S. Bandyopadhyay, S. Bhattacharjee and D. Sen,
Driven quantum many-body systems and out-of-equilibrium topology,
J. Phys.: Condens. Matter \textbf{33}, 393001 (2021).

\bibitem{FTPRev08} A. de la Torre, D. M. Kennes, M. Claassen, S.
Gerber, J. W. McIver, and M. A. Sentef, Colloquium: Nonthermal pathways
to ultrafast control in quantum materials, Rev. Mod. Phys. \textbf{93},
041002 (2021).

\bibitem{FTPRev09} L. Zhou and D.-J. Zhang, Non-Hermitian Floquet
topological matter---A review, Entropy \textbf{25}, 1401 (2023).

\bibitem{FTPRev10} F. Zhan, R. Chen, Z. Ning, D.-S. Ma, Z. Wang,
D.-H. Xu, and R. Wang, Perspective: Floquet engineering topological
states from effective models towards realistic materials, Quantum
Front. \textbf{3}, 21 (2024).

\bibitem{FloLarge01} D. Y. H. Ho and J. Gong, Quantized adiabatic
transport in momentum space, Phys. Rev. Lett. \textbf{109}, 010601
(2012).

\bibitem{FloLarge02} Q.-J. Tong, J.-H. An, J. Gong, H.-G. Luo, and
C. H. Oh, Generating many Majorana modes via periodic driving: A superconductor
model, Phys. Rev. B \textbf{87}, 201109(R) (2013).

\bibitem{FloLarge03} A. Kundu, H. A. Fertig, and B.
Seradjeh, Effective Theory of Floquet Topological Transitions, Phys.
Rev. Lett. \textbf{113}, 236803 (2014).

\bibitem{FloLarge04} T.-S. Xiong, J. Gong, and J.-H. An, Towards
large-Chern-number topological phases by periodic quenching, Phys.
Rev. B \textbf{93}, 184306 (2016).

\bibitem{FloLarge05} L. Zhou and J. Gong, Recipe for creating an
arbitrary number of Floquet chiral edge states, Phys. Rev. B \textbf{97},
245430 (2018).

\bibitem{FloLarge06} K. Yang, S. Xu, L. Zhou, Z. Zhao, T. Xie, Z.
Ding, W. Ma, J. Gong, F. Shi, and J. Du, Observation of Floquet topological
phases with large Chern numbers, Phys. Rev. B \textbf{106}, 184106
(2022).

\bibitem{AFTP01} L. Jiang, T. Kitagawa, J. Alicea, A. R. Akhmerov,
D. Pekker, G. Refael, J. I. Cirac, E. Demler, M. D. Lukin, and P.
Zoller, Majorana fermions in equilibrium and in driven cold-atom quantum
wires, Phys. Rev. Lett. \textbf{106}, 220402 (2011).

\bibitem{AFTP02} M. S. Rudner, N. H. Lindner, E. Berg, and M. Levin,
Anomalous edge states and the bulk-edge correspondence for periodically
driven two-dimensional systems, Phys. Rev. X \textbf{3}, 031005 (2013).

\bibitem{AFTP03} L. Zhou, C. Chen, and J. Gong, Floquet semimetal
with Floquet-band holonomy, Phys. Rev. B \textbf{94}, 075443 (2016).

\bibitem{AFTP04} L. J. Maczewsky, J. M. Zeuner, S. Nolte, and A.
Szameit, Observation of photonic anomalous Floquet topological insulators,
Nat. Commun. \textbf{8}, 13756 (2017).

\bibitem{AFTP05} M. Rodriguez-Vega and B. Seradjeh, Universal Fluctuations
of Floquet Topological Invariants at Low Frequencies, Phys. Rev. Lett.
\textbf{121}, 036402 (2018).

\bibitem{AFTP06} K. Wintersperger, C. Braun, F. N. \"Unal, A. Eckardt,
M. D. Liberto, N. Goldman, I. Bloch, and M. Aidelsburger, Realization
of an anomalous Floquet topological system with ultracold atoms, Nat.
Phys. \textbf{16}, 1058 (2020).

\bibitem{AFTP07} M. Li, C. Li, L. Yan, Q. Li, Q. Gong, and Y. Li,
Fractal photonic anomalous Floquet topological insulators to generate
multiple quantum chiral edge states, Light Sci. Appl. \textbf{12},
262 (2023).

\bibitem{AFTP08} M. Ghuneim and R. W. Bomantara, Anomalous topological edge modes in a periodically driven trimer lattice,
Phys. Rev. B {\bf 111}, 195424 (2025).

\bibitem{FgSPT01} L. Zhou, J. Gong, and X.-J. Yu, Topological edge
states at Floquet quantum criticality, Commun. Phys. \textbf{8}, 214
(2025).

\bibitem{FgSPT02} G. Cardoso, H.-C. Yeh, L. Korneev, A. G. Abanov,
and A. Mitra, Gapless Floquet topology, Phys. Rev. B \textbf{111},
125162 (2025).

\bibitem{FgSPT03} L. Zhou, R. Wang, and J. Pan, Gapless higher-order
topology and corner states in Floquet systems, Phys. Rev. Res. \textbf{7},
023079 (2025).

\bibitem{FloClass01} T. Kitagawa, E. Berg, M. Rudner, and E. Demler,
Topological characterization of periodically driven quantum systems,
Phys. Rev. B \textbf{82}, 235114 (2010).

\bibitem{FloClass02} F. Nathan and M. S. Rudner, Topological singularities
and the general classification of Floquet-Bloch systems, New J. Phys.
\textbf{17}, 125014 (2015).

\bibitem{FloClass03} S. Yao, Z. Yan, and Z. Wang, Topological invariants
of Floquet systems: General formulation, special properties, and Floquet
topological defects, Phys. Rev. B \textbf{96}, 195303 (2017).

\bibitem{FloquetPT} R. Roy and F. Harper, Periodic table for Floquet
topological insulators, Phys. Rev. B \textbf{96}, 155118 (2017).

\bibitem{FloCry01} T. Morimoto, H. C. Po, and A. Vishwanath, Floquet
topological phases protected by time glide symmetry, Phys. Rev. B
\textbf{95}, 195155 (2017).

\bibitem{FloCry02} S. Xu and C. Wu, Space-Time Crystal and Space-Time
Group, Phys. Rev. Lett. \textbf{120}, 096401 (2018).

\bibitem{FloCry03} S. Franca, J. van den Brink, and I. C. Fulga,
An anomalous higher-order topological insulator, Phys. Rev. B \textbf{98},
201114(R) (2018).

\bibitem{FloCry04} R. W. Bomantara, L. Zhou, J. Pan, and J. Gong,
Coupled-wire construction of static and Floquet second-order topological
insulators, Phys. Rev. B \textbf{99}, 045441 (2019).

\bibitem{FloCry05} M. Rodriguez-Vega, A. Kumar, and B. Seradjeh,
Higher-order Floquet topological phases with corner and bulk bound
states, Phys. Rev. B \textbf{100}, 085138 (2019).

\bibitem{FloCry06} Y. Peng and G. Refael, Floquet Second-Order Topological
Insulators from Nonsymmorphic Space-Time Symmetries, Phys. Rev. Lett.
\textbf{123}, 016806 (2019).

\bibitem{FloCry07} K. Ladovrechis and I. C. Fulga, Anomalous Floquet
topological crystalline insulators, Phys. Rev. B \textbf{99}, 195426
(2019).

\bibitem{FloCry08} R. Seshadri, A. Dutta, and D. Sen, Generating
a second-order topological insulator with multiple corner states by
periodic driving, Phys. Rev. B \textbf{100}, 115403 (2019).

\bibitem{FloCry09} K. Plekhanov, M. Thakurathi, D. Loss, and J. Klinovaja,
Floquet second-order topological superconductor driven via ferromagnetic
resonance, Phys. Rev. Res. \textbf{1}, 032013(R) (2019).

\bibitem{FloCry092} T. Nag, V. Jurici{\v c}i\'c and B. Roy, Out of equilibrium higher-order topological insulator: Floquet engineering and quench dynamics,
Phys. Rev. Res. {\bf 1}, 032045(R) (2019).

\bibitem{FloCry10} S. Chaudhary, A. Haim, Y. Peng, and G. Refael,
Phonon-induced Floquet topological phases protected by space-time
symmetries, Phys. Rev. Res. \textbf{2}, 043431 (2020).

\bibitem{FloCry11} A. K. Ghosh, G. C. Paul, and A. Saha, Higher order
topological insulator via periodic driving, Phys. Rev. B \textbf{101},
235403 (2020).

\bibitem{FloCry12} H. Hu, B. Huang, E. Zhao, and W. V. Liu, Dynamical
Singularities of Floquet Higher-Order Topological Insulators, Phys.
Rev. Lett. \textbf{124}, 057001 (2020).

\bibitem{FloCry13} B. Huang and W. V. Liu, Floquet Higher-Order Topological
Insulators with Anomalous Dynamical Polarization, Phys. Rev. Lett.
\textbf{124}, 216601 (2020).

\bibitem{FloCry14} Y. Peng, Floquet higher-order topological insulators
and superconductors with space-time symmetries, Phys. Rev. Res. \textbf{2},
013124 (2020).

\bibitem{FloCry15} J. Pan and L. Zhou, Non-Hermitian Floquet second
order topological insulators in periodically quenched lattices, Phys.
Rev. B \textbf{102}, 094305 (2020).

\bibitem{FloCry16} T. Nag, V. Juri{\v c}i\'c,
and B. Roy, Hierarchy of higher-order Floquet topological phases in
three dimensions, Phys. Rev. B \textbf{103}, 115308 (2021).

\bibitem{FloCry17} A. K. Ghosh, T. Nag, and A. Saha, Floquet generation
of a second-order topological superconductor, Phys. Rev. B \textbf{103},
045424 (2021).

\bibitem{FloCry18} R.-X. Zhang and Z.-C. Yang, Tunable fragile topology
in Floquet systems, Phys. Rev. B \textbf{103}, L121115 (2021).

\bibitem{FloCry19} W. Zhu, Y. D. Chong, and J. Gong, Floquet higher-order
topological insulator in a periodically driven bipartite lattice,
Phys. Rev. B \textbf{103}, L041402 (2021).

\bibitem{FloCry20} H. Chen and W. V. Liu, Intertwined space-time
symmetry, orbital magnetism, and dynamical Berry connection in a circularly
shaken optical lattice, Phys. Rev. A \textbf{104}, 013308 (2021).

\bibitem{FloCry21} J. Yu, R.-X. Zhang, and Z.-D. Song, Dynamical
symmetry indicators for Floquet crystals, Nat. Commun. \textbf{12},
5985 (2021).

\bibitem{FloCry22} W. Zhu, H. Xue, J. Gong, Y. Chong, and B. Zhang,
Time-periodic corner states from Floquet higher-order topology, Nat.
Commun. \textbf{13}, 11 (2022).

\bibitem{FloCry23} Z. Cheng, R. W. Bomantara, H. Xue, W. Zhu, J.
Gong, and B. Zhang, Observation of $\pi/2$ Modes in an Acoustic Floquet
System, Phys. Rev. Lett. \textbf{129}, 254301 (2022).

\bibitem{MTI0} Y. Deng and Y. Jing, Acoustic Crystals with a M\"obius
Twist, Physics \textbf{15}, 36 (2022).

\bibitem{TenfoldNJP} S. Ryu, A. P. Schnyder, A. Furusaki, and A.
W. W. Ludwig, Topological insulators and superconductors: tenfold
way and dimensional hierarchy, New J. Phys. \textbf{12}, 065010 (2010).

\bibitem{Sym1} L. Michel and J. Zak, Connectivity of energy bands
in crystals, Phys. Rev. B \textbf{59}, 5998 (1999).

\bibitem{MTI01} F. Zhang and C. L. Kane, Anomalous topological pumps
and fractional Josephson effects, Phys. Rev. B \textbf{90}, 020501(R)
(2014).

\bibitem{MTI02} S. M. Young and C. L. Kane, Dirac Semimetals in Two
Dimensions, Phys. Rev. Lett. \textbf{115}, 126803 (2015).

\bibitem{MTI03} K. Shiozaki, M. Sato, and K. Gomi, $\mathbb{Z}_{2}$
topology in nonsymmorphic crystalline insulators: M\"obius twist in
surface states, Phys. Rev. B \textbf{91}, 155120 (2015).

\bibitem{MTI04} Y. X. Zhao and A. P. Schnyder, Nonsymmorphic symmetry-required
band crossings in topological semimetals, Phys. Rev. B \textbf{94},
195109 (2016).

\bibitem{MTI05} P.-Y. Chang, O. Erten, and P. Coleman, M\"obius Kondo
insulators, Nat. Phys. \textbf{13}, 794 (2017).

\bibitem{MTI06} R.-X. Zhang, F. Wu, and S. D. Sarma, M\"obius insulator
and higher-order topology in MnBi$_{2n}$Te$_{3n+1}$, Phys. Rev.
Lett. \textbf{124}, 136407 (2020).

\bibitem{MTI07} Y. X. Zhao, Y.-X. Huang, and S. A. Yang, $\mathbb{Z}_{2}$-projective
translational symmetry protected topological phases, Phys. Rev. B
\textbf{102}, 161117(R) (2020).

\bibitem{MTI08} L. B. Shao, Q. Liu, R. Xiao, S. A. Yang,
and Y. X. Zhao, Gauge-Field Extended $k\cdot p$ Method
and Novel Topological Phases, Phys. Rev. Lett. \textbf{127}, 076401
(2021).

\bibitem{MTI09} Y. Yang, H. C. Po, V. Liu, J. D. Joannopoulos, L.
Fu, and M. Solja{\v c}i\'c, Non-Abelian
nonsymmorphic chiral symmetries, Phys. Rev. B \textbf{106}, L161108
(2022).

\bibitem{MTI10} H. Xue, Z. Wang, Y.-X. Huang, Z. Cheng, L. Yu, Y. X.
Foo, Y. X. Zhao, S. A. Yang, and B. Zhang, Projectively
Enriched Symmetry and Topology in Acoustic Crystals, Phys. Rev. Lett.
\textbf{128}, 116802 (2022).

\bibitem{MTI11} T. Li, J. Du, Q. Zhang, Y. Li, X. Fan, F. Zhang,
and C. Qiu, Acoustic M\"obius Insulators from Projective Symmetry, Phys.
Rev. Lett. \textbf{128}, 116803 (2022).

\bibitem{MTI12} Y.-X. Huang, Z. Y. Chen, X. Feng, S. A. Yang, and
Y. X. Zhao, Periodic Clifford symmetry algebras on flux lattices,
Phys. Rev. B \textbf{106}, 125102 (2022).

\bibitem{MTI13} Z.-M. Yu, Z. Zhang, G.-B. Liu, W. Wu, X.-P. Li, R.-W.
Zhang, S. A. Yang, and Y. Yao, Encyclopedia of emergent particles
in three-dimensional crystals, Sci. Bull. \textbf{67}, 375 (2022).

\bibitem{MTI14} Z. Liu, G. Wei, H. Wu, and J.-J. Xiao, M$\ddot{{\rm o}}$bius
edge band and Weyl-like semimetal flat-band in topological photonic
waveguide array by synthetic gauge flux, Nanophotonics \textbf{12},
3481 (2023).

\bibitem{MTI15} C. Jiang, Y. Song, X. Li, P. Lu, and S. Ke, Photonic
M\"obius topological insulator from projective symmetry in multiorbital
waveguides, Opt. Lett. \textbf{48}, 2337 (2023).

\bibitem{MTI16} S. Bao, J. Chang, J. Wu, and Z. Xu, Circuit realization
of M\"obius insulators, Phys. Rev. A \textbf{108}, 013508 (2023).

\bibitem{MTI17} F. Gao, Y.-G. Peng, Q.-L. Sun, X. Xiang, C. Zheng,
and X.-F. Zhu, Topological acoustics with orbital-dependent gauge
fields, Phys. Rev. App. \textbf{20}, 064036 (2023).

\bibitem{MTI18} Z. Y. Chen, Z. Zhang, S. A. Yang, and Y. X. Zhao,
Classification of time-reversal-invariant crystals with gauge structures,
Nat. Commun. \textbf{14}, 743 (2023).

\bibitem{MTI19} Y. Liu, C. Jiang, W. Wen, Y. Song, X. Li, P. Lu,
and S. Ke, Topological phases in photonic microring lattices with
projective symmetry, Phys. Rev. A \textbf{109}, 013516 (2024).

\bibitem{MTI20} Q. Wang, Z. Fu, L. Ye, H. He, W. Deng, J. Lu, M.
Ke, and Z. Liu, Non-Hermitian acoustic M\"obius insulator, Phys. Rev.
B \textbf{111}, L100101 (2025).

\bibitem{MTI21} S.-N. Liang, J.-L. Xie, C. He, S.-Y. Yu, and Y.-F.
Chen, Manipulating M\"obius edge states in elastic wave phononic crystals,
Phys. Rev. B \textbf{111}, 184103 (2025).

\bibitem{MTI22} Y. Deng and Y. Jing, Acoustic Crystals with a M\"obius
Twist, Physics \textbf{15}, 36 (2022).

\bibitem{SSH} W. P. Su, J. R. Schrieffer, and A. J. Heeger, Solitons
in Polyacetylene, Phys. Rev. Lett. \textbf{42}, 1698 (1979).

\bibitem{STF01} J. K. Asb\'oth and H. Obuse, Bulk-boundary
correspondence for chiral symmetric quantum walks, Phys. Rev. B \textbf{88},
121406(R) (2013).

\bibitem{STF02} D. Y. H. Ho and J. Gong, Topological effects in chiral
symmetric driven systems, Phys. Rev. B \textbf{90}, 195419 (2014).

\bibitem{STF03} L. Zhou and J. Gong, Floquet topological phases in
a spin-$1/2$ double kicked rotor, Phys. Rev. A \textbf{97}, 063603
(2018).

\bibitem{BBH} W. A. Benalcazar, B. A. Bernevig, and T. L. Hughes,
Quantized electric multipole insulators, Science \textbf{357}, 61
(2017).

\bibitem{ESEE01} L. Zhou, Entanglement spectrum and entropy in Floquet
topological matter, Phys. Rev. Res. \textbf{4}, 043164 (2022).

\bibitem{gSPT0} R. Verresen, N. G. Jones, and F. Pollmann, Topology
and Edge Modes in Quantum Critical Chains, Phys. Rev. Lett. \textbf{120},
057001 (2018).

\bibitem{PMCOUPLE} Y. Qi, C. Qiu, M. Xiao, H. He, M. Ke, and Z. Liu, Acoustic Realization of Quadrupole Topological Insulators, Phys. Rev. Lett. \textbf{124}, 206601 (2020).

\end{thebibliography}
\end{document}